\documentclass{article}
\usepackage[utf8]{inputenc}
\usepackage[T1]{fontenc}
\usepackage{graphicx}
\usepackage{amsmath}
\usepackage{amssymb}
\usepackage{overcite}
\usepackage{xfrac}
\usepackage{booktabs}
\usepackage[font=small,labelfont=bf]{caption}

\newcommand{\onlinecite}[1]{\hspace{-1 ex} \nocite{#1}\citenum{#1}}

\title{Gaussian Process-Based Refinement of Dispersion Corrections}
\author{Jonny Proppe,$^{a,b,\ast,\dagger}$ Stefan Gugler$^{b,\dagger}$ and Markus Reiher$^{b,}$\thanks{\textsf{corresponding authors: jonny.proppe@utoronto.ca, markus.reiher@phys.chem.ethz.ch}}}
\date{\today}

\begin{document}

\maketitle 

\vspace*{-0.9cm}\begin{center}
$^a$Departments of Chemistry and Computer Science, University of Toronto, Toronto, Ontario, Canada. \\
$^b$Laboratory of Physical Chemistry, ETH Zurich, Vladimir-Prelog-Weg 2, 8093 Zurich, Switzerland. \\
$^\dagger$ These authors contributed equally.
\end{center}

\begin{abstract}
    \noindent We employ Gaussian process (GP) regression to adjust for systematic errors in D3-type dispersion corrections.
    We refer to the associated, statistically improved model as D3-GP.
    It is trained on differences between interaction energies obtained from PBE-D3(BJ)/ma-def2-QZVPP and DLPNO-CCSD(T)/CBS calculations.
    We generated a data set containing interaction energies for 1,248 molecular dimers, which resemble the dispersion-dominated systems contained in the S66 data set.
    Our systems do not only represent equilibrium structures, but also dimers with various relative orientations and conformations at both shorter and longer distances.
    A reparametrization of the D3(BJ) model based on 66 of these dimers suggests that two of its three empirical parameters, $a_1$ and $s_8$, are zero, whereas ${a_2 = 5.6841}$~bohr.
    For the remaining 1,182 dimers, we find that this new set of parameters is superior to all previously published D3(BJ) parameter sets.
    To train our D3-GP model, we engineered two different vectorial representations of (supra-)molecular systems, both derived from the matrix of atom-pairwise D3(BJ) interaction terms:
    (a) a distance-resolved interaction energy histogram, \textsf{histD3(BJ)}, and (b) eigenvalues of the interaction matrix ordered according to their decreasing absolute value, \textsf{eigD3(BJ)}.
    Hence, the GP learns a mapping from D3(BJ) information only, which renders D3-GP-type dispersion corrections comparable to those obtained with the original D3 approach.
    They improve systematically if the underlying training set is selected carefully.
    Here, we harness the prediction variance obtained from GP regression to select optimal training sets in an automated fashion.
    The larger the variance, the more information the corresponding data point may add to the training set.
    For a given set of molecular systems, variance-based sampling can approximately determine the smallest subset being subjected to reference calculations such that all dispersion corrections for the remaining systems fall below a predefined accuracy threshold.
    To render the entire D3-GP workflow as efficient as possible, we present an improvement over our variance-based, sequential active-learning scheme [\textit{J. Chem. Theory Comput.} \textbf{2018}, \textit{14}, 5238].
    Our refined learning algorithm selects multiple (instead of single) systems which can be subjected to reference calculations simultaneously. 
    We refer to the underlying selection strategy as batch-wise variance-based sampling (BVS).
	BVS-guided active learning is an essential component of our D3-GP workflow, which is implemented in a black-box fashion.
	Once provided with reference data for new molecular systems, the underlying GP model automatically learns to adapt to these and similar systems.
	This approach leads overall to a self-improving model (D3-GP) that predicts system-focused and GP-refined D3-type dispersion corrections for any given system of reference data.     
	   
\end{abstract}

\newpage

\section{Introduction}
	The inability of (semi)local exchange--correlation functionals to capture long-range electron correlation and, hence, London dispersion interactions, was demonstrated several times in the past two decades.\cite{kristyan1994, perez-jorda1995, ruzsinszky2005, johnson2006a, wodrich2006, wodrich2007}
	Three major groups of approaches were developed to account for dispersion interactions in density functional theory (DFT): effective one-electron potentials, nonlocal density functionals, and semiclassical dispersion corrections.
	Important examples of the first two groups are the atom-centered nonlocal potentials by Röthlisberger and co-workers,\cite{lilienfeld2004} and the Vydrov--Van Voorhis nonlocal density functional.\cite{vydrov2010}
	A discussion of these nonadditive approaches is beyond the scope of this paper, and we refer the reader to Refs.\,\onlinecite{grimme2016}~and~\onlinecite{hermann2017} for general reviews addressing dispersion corrections for electronic-structure methods.
	Here, we consider semiclassical dispersion corrections, which are simply added to the electronic energy.
	The computational efficiency resulting from this additive character indicates why the semiclassical approach is the most widely used approach to take dispersion interactions into account.
	It has been proposed as early as the 1970s.\cite{cohen1974, ahlrichs1977} 
	Significant advances were achieved in the 2000s and 2010s by Becke and Johnson,\cite{johnson2005, johnson2006, becke2007} Tkatchenko and Scheffler,\cite{tkatchenko2009, tkatchenko2012} Steinmann and Corminboeuf,\cite{steinmann2010, steinmann2011} Sato and Nakai,\cite{sato2009, sato2010} and 
	Grimme and co-workers\cite{grimme2004, grimme2006, grimme2010, grimme2011, caldeweyher2017, caldeweyher2019} (D$x$-type dispersion corrections with $x=1,2,3,4$).
        
    The popularity of D$x$-type dispersion correction models\cite{grimme2004, grimme2006, grimme2010, grimme2011, caldeweyher2017, caldeweyher2019} may be attributed to (i) their good performance across the periodic table, (ii) their availability for many mean-field electronic-structure approximations, (iii) their incorporation in various quantum-chemical computer programs, and (iv) their very low computational complexity.
    Originally optimized for equilibrium structures, Smith~et~al.\cite{smith2016}~showed that a mere reparametrization of the most popular D$x$ models, D3(0)\cite{grimme2010} (with the original, zero-damping function) and D3(BJ)\cite{grimme2011} (with the Becke--Johnson damping function), also allows for an accurate description of noncovalent interactions across configuration space.
    Further improvements over the original D3 approach have recently been released with D4.\cite{caldeweyher2017, caldeweyher2019}
    Compared to D3-type dispersion corrections, the estimation of dispersion coefficients is not only based on the molecular structure, but also involves partial charges obtained from a classical model (default) or, alternatively, from electronic-structure calculations.
    In the latter case, the dependence on the electron density renders the D4 approach more similar to other semiclassical dispersion correction schemes that also incorporate electronic-structure information.\cite{johnson2005, johnson2006, becke2007, tkatchenko2009, tkatchenko2012, steinmann2010, steinmann2011, sato2009, sato2010}
    
    Predictions derived from truncated physical models (such as those discussed in this work) can be expected to be biased in a nontrivial way with respect to the underlying model variables.\cite{simm2017}
    To identify such systematic errors, one could, in principle, carry out reference measurements or reference calculations in a case-by-case fashion.
    However, this approach is unfeasible and would render any prediction unnecessary. 
    One would rather prefer a measure that informs about the reliability of a prediction, here, an a posteriori correction of the approximate interaction energy of a given (supra-)molecular system.
    The variance of a system-specific dispersion correction obtained from a statistical analysis is one possible measure. 
    Recently, we showed how to estimate such variances by sampling the uncertainty in the D3(BJ) parameters with bootstrapping.\cite{weymuth2018} 
    However, parameter uncertainty may explain only a small fraction of the total prediction error, which is usually dominated by the inadequacy of the model itself.\cite{pernot2015, proppe2017}
    
    Machine learning\cite{bishop2006} offers a remedy for alleviating model inadequacy as the performance of machine-learned models improves with data.
    Only a few studies have been published to date which probe the possibilities of machine learning for noncovalent-interaction research.
    Mezei~et~al.\cite{mezei2019}~applied kernel ridge regression\cite{rupp2015} in combination with many-body descriptors\cite{faber2018} to learn corrections for interaction energies obtained from DFT calculations.
    Instead of learning corrections to the interaction energy directly, they learned differences to CCSD(T) atomization energies to also improve on intramolecular interactions.
    McGibbon et~al.\cite{mcgibbon2017}~learned corrections for interaction energies obtained from second-order M{\o}ller--Plesset perturbation theory calculations.
    They applied neural-network regression\cite{bishop2006} on descriptors that were derived from various electronic-structure approximations, e.g., density matrix overlaps and components of noncovalent interactions (dispersion, electrostatics, exchange, and induction).
    CCSD(T) interaction energies for 200,000 molecules across the periodic table were employed to train their neural network.
    They estimated system-specific prediction uncertainties by generating 1,000 instances of their neural network via dropout.\cite{srivastava2014}
    Gao et~al.\cite{gao2016}~used partial least-squares regression\cite{hastie2009} to select descriptors such as electronic energies or HOMO--LUMO gaps, which were obtained from DFT calculations.
    A neural network was trained on these descriptors to learn corrections to DFT interaction energies with respect to a CCSD(T) reference.
    An alternative approach to learning corrections for interaction energies was proposed by Bereau~et~al.\cite{bereau2018}
    They applied kernel ridge regression to estimate local environment-dependent properties such as electrostatic multipole coefficients and polarizabilities.
    These machine-learned quantities were then fed into classical intermolecular potentials to improve the prediction of many-body dispersion and other types of noncovalent interactions.
    Machine-learned models for electron densities\cite{brockherde2017,bogojeski2018} have been adapted for highly scalable models\cite{grisafi2019} and to gain insights into noncovalent interaction\cite{fabrizio2019} by Corminboeuf, Ceriotti, and co-workers.
    Machine learning usually is a data-extensive endeavor as the corresponding prediction models potentially improve with an increasing number of benchmark data.
    However, the calculation of benchmark interaction energies constitutes a computational bottleneck.
    This limitation offers the incentive to generate informative training sets, which contain a minimum number of entries to achieve a predefined accuracy over a given domain of application, i.e., the chemical and configuration space covered by the set of systems at hand.
    It is not sufficient to sample the existing set of systems uniformly, as the number of systems that are required to achieve that accuracy is not known a priori.
    Therefore, a dynamic extension of the training set is necessary, e.g., by means of active learning.\cite{settles2012}
    In active-learning approaches, predictions learned from a given training set are harnessed to determine a batch of new systems for which benchmarks are required.
    The training set will be extended by these benchmarks to optimally improve the underlying prediction model for those systems that are not yet contained in the training set.
    
    Regarding research in computational chemistry, controlling the number of benchmarks is particularly important in data-extensive scenarios such as molecular-dynamics simulations and chemical-space exploration.
    In such scenarios, not even the set of systems is known a priori.
    Botu and Ramprasad\cite{botu2015} compared similarities between molecular fingerprints\cite{bartok2013} (vectorial representations of molecules) to actively learn energies and forces for molecular-dynamics simulations.
    If a new configuration along the trajectory is too dissimilar to the training set, electronic-structure calculations are performed to update the machine-learned force field.
    Another approach was described by Behler\cite{behler2015} to select training data for learning high-dimensional neural-network potentials.
    Several neural networks with different architectures but similar performances are employed to predict potential energies for a given trajectory obtained from one of these networks.
    Due to the choice of multiple neural networks, a prediction variance can be estimated for each configuration along the trajectory.
    High variances indicate a disagreement between the different neural-network models.
    Therefore, high-variance configurations are subjected to electronic-structure calculations to update these models.
    This approach is sometimes denoted ``Query by Committee''.\cite{smith2018}
    Active-learning approaches have also been applied to accelerate chemical-space exploration (based on Thompson sampling\cite{hernandez2017}) and for the prediction of molecular properties (based on the D-optimality criterion\cite{gubaev2018}).
    The seminal work by Cs\'{a}nyi et al. can be considered an early predecessor to the active-learning approach developed here.\cite{csanyi2004}
    
    In this work, we adjust for systematic errors in D3(BJ) dispersion corrections by exploiting the capabilities of Gaussian process (GP) regression,\cite{rasmussen2006} the Bayesian analog to kernel ridge regression.
    We refer to the resulting prediction model as D3(BJ)-GP.
    The only input it requires is the structure of a (supra-)molecular system, i.e., its atomic positions and corresponding proton numbers, which are transformed to linear combinations of atom-pairwise D3(BJ) interaction terms.
    This way, the computational efficiency of the original D3(BJ) approach can be preserved to a high degree.
    Due to the Bayesian nature of GP regression, D3(BJ)-GP dispersion corrections are being sampled from a posterior probability distribution.
    The variance of this distribution can be considered a reliability measure.
    If the variance is small with respect to a user-defined tolerance, the mean of the posterior distribution can be considered a reliable dispersion correction.
    If the variance is too large, however, the user is advised to perform reference electronic-structure calculations in order to generate a new benchmark, extend the training set, and update the GP posterior distribution.
    Here, we harness the posterior variance as a simple selection criterion to actively learn dispersion corrections.
    Hence, the training set employed to optimize the D3(BJ)-GP model is dynamic, which constitutes an essential novelty to the active research field of dispersion-corrected electronic-structure methods.\cite{grimme2016, hermann2017}
    As a result, a system-focused model is obtained that by virtue of automated reference calculations can improve itself in a rolling fashion.
    Naturally, a meta machine-learning model may be applied to extract transferable knowledge from sets of such system-focused self-improving models.
    Furthermore, we present an extension to our original, sequential formulation of variance-based active learning,\cite{simm2018} which allows us to generate multiple benchmarks in parallel.
    We refer to our strategy of selecting multiple molecular systems prior to benchmark generation as batch-wise variance-based sampling (BVS).
    BVS is particularly interesting for data-extensive scenarios as mentioned above.
    
    This paper is structured as follows:
    A concise introduction to GP regression and related topics (Bayesian optimization and BVS) is provided in Section~\ref{sec:gpr}.
    In Section~\ref{sec:methods}, we introduce the data sets studied in this work, provide details on electronic-structure calculations, and describe the settings employed for GP regression and subsequent analyses.
    The results of this study are presented and discussed in Section~\ref{sec:results}.
    We report on a problem-specific D3(BJ) reparametrization, the performance of the D3(BJ)-GP model based on a fixed training set, and the improvement of the D3(BJ)-GP performance resulting from BVS-guided active learning.
  
\section{Gaussian Process Regression}
\label{sec:gpr}

    The GP posterior distribution of the \textit{target} variable $y$, which may be a physical property, is Gaussian by definition and can, therefore, be fully specified by both its mean and its variance.
    As with any other regression method, GP regression is applied to learn a mapping from a domain $\{x_1,\cdots,x_M\}$ of $M$ continuous input variables (\textit{features}) to $y$.
    The features may be descriptors of a physical system.
    Given $N$ observations of the target variable, $\mathbf{y} = (y_1,\cdots,y_N)^\top$, mean and variance mapped from any feature vector $\mathbf{x}_\ast = (x_{\ast,1},\cdots,x_{\ast,M})$ read
    
\begin{equation}
\label{eq:mean}
	\mathbb{E}\big[\hat{y}(\mathbf{x}_\ast)\big] \equiv \hat{\mu}_y(\mathbf{x}_\ast) = \mathbf{k}(\mathbf{x}_\ast,\mathbf{X}) \Big( \mathbf{K}(\mathbf{X},\mathbf{X}) + \alpha_0\mathbf{I} \Big)^{-1} \mathbf{y} \
\end{equation}

and

\begin{equation}
\label{eq:var}
	\mathbb{V}\big[\hat{y}(\mathbf{x}_\ast)\big] \equiv \hat{\sigma}_y^2(\mathbf{x}_\ast) = k(\mathbf{x}_\ast,\mathbf{x}_\ast) -  \mathbf{k}(\mathbf{x}_\ast,\mathbf{X}) \Big( \mathbf{K}(\mathbf{X},\mathbf{X}) + \alpha_0\mathbf{I} \Big)^{-1} \mathbf{k}(\mathbf{X},\mathbf{x}_\ast) \ ,
\end{equation}

    respectively.
	The rows of the design matrix $\mathbf{X}$ represent the $N$ feature vectors $\{ \mathbf{x}_n = (x_{n,1},\cdots,x_{n,M})\}$ of the observed (training) set.
    The kernel $k(\mathbf{x}_i,\mathbf{x}_j)$ is a measure of similarity between feature vectors $\mathbf{x}_i$ and $\mathbf{x}_j$, and the row vector $\mathbf{k}(\mathbf{x}_i,\mathbf{X})$ as well as the matrix $\mathbf{K}(\mathbf{X},\mathbf{X})$ are the similarity between one feature vector, $\mathbf{x}_i$, or a matrix of features, $\mathbf{X}$, with the design matrix $\mathbf{X}$. 
	The hyperparameter $\alpha_0$, multiplied with the $(N \times N)$-dimensional identity matrix $\mathbf{I}$, represents the constant (homoscedastic) noise of each observation.
	We refer the reader to Ref.~\citeonline{rasmussen2006} for a comprehensive introduction to the topic.
	
\subsection{Bayesian Optimization}

    The mean of the GP posterior distribution, $\hat{\mu}_y(\mathbf{x})$, represents the best approximation to the target function $y(\mathbf{x})$.
    As evaluations (observations) of $\hat{\mu}_y(\mathbf{x})$ are much faster than evaluations of $y(\mathbf{x})$, the former can also be used to accelerate searches on $y(\mathbf{x})$ if one is interested in optimization tasks.
    In such cases, $y(\mathbf{x})$ serves as an objective function.
    To ensure finding the global minimum (or other extrema) of $y(\mathbf{x})$, we target an optimal balance between probing its already known, promising regions and exploring its yet unknown regions.
    Bayesian optimization\cite{shahriari2016} achieves this goal by combining mean and variance of the GP posterior distribution, $\hat{\mu}_y(\mathbf{x})$ and $\hat{\sigma}^2_y(\mathbf{x})$, to construct an acquisition function ${u(\mathbf{x}) = f\left(\hat{\mu}_y(\mathbf{x}),\hat{\sigma}^2_y(\mathbf{x})\right)}$, which can be interpreted as a purposely biased objective function.
    The true objective $y(\mathbf{x}_\ast)$ is evaluated at the current minimum $\mathbf{x}_\ast$ of $u(\mathbf{x})$, which, in turn, is used to update the posterior distribution.

\subsection{Variance-Based Active Learning}
    \label{sec:var-based}
    Consider a training set $\mathcal{T}^{N_y}_{N_\mathbf{x}}$ with $N_\mathbf{x}$ feature vectors and $N_y$ observations, where $N_\mathbf{x} = N_y$, and a query set $\mathcal{Q}_{M_\mathbf{x}}$ with $M_\mathbf{x}$ feature vectors.
	We seek to add the smallest subset of $\mathcal{Q}_{M_\mathbf{x}}$ to $\mathcal{T}^{N_y}_{N_\mathbf{x}}$ such that \textit{every} feature vector of the remaining query set is assigned a posterior standard deviation $\hat{\sigma}_y(\mathbf{x}_\ast)$ that is below a predefined accuracy threshold $t$.
	Promoting the query vector with the highest posterior variance to a training vector is a straightforward strategy in this respect.
	Note that the hyperparameters are a function of the training set defined by $\mathbf{X}$ and $\mathbf{y}$ and, hence, the variance $\hat{\sigma}_y^2$ as defined in Eq.~(\ref{eq:var}) depends implicitly on $\mathbf{y}$.
	As a consequence, updating $\hat{\sigma}_y^2$ requires an observation after \textit{each} promotion of a query vector.
	This \textit{sequential} approach\cite{simm2018} to making observations constitutes a critical computational bottleneck.
	
	To render our approach more efficient, we determine a batch size $L$ that reflects the number of observations we can actually make in \textit{parallel}.
    We determine the query vector with the highest posterior variance, $\hat{\sigma}_y^2$, and add its feature vector to $\mathbf{X}$, resulting in $\mathcal{Q}_{M_\mathbf{x}-1}$ and $\mathcal{T}_{N_\mathbf{x}+1}^{N_y}$.
	Subsequently, we update $\hat{\sigma}_y^2$ \textit{without} a preceding hyperparameter optimization.
	As we temporarily fix the hyperparameters, $\hat{\sigma}_y^2$ does no longer depend on $\mathbf{y}$ and, therefore, we can update the variance without making observations.
	We perform this subalgorithm (query vector promotion followed by a fixed-hyperparameter variance update) at most $L$ times to obtain $\mathcal{Q}_{M_\mathbf{x}-l}$ and $\mathcal{T}_{N_\mathbf{x}+l}^{N_y}$, where $l \in \{ 1,\cdots,L \}$.
	The subalgorithm will stop if either $\hat{\sigma}_y(\mathbf{x}_m) < t ~ \forall \mathbf{x}_m \in \mathcal{Q}_{M_\mathbf{x}-l}$ or $l = L$.
	After adding a batch of $l$ feature vectors from the query set to the training set, we make $l$ observations in parallel to obtain $\mathcal{T}_{N_\mathbf{x}+l}^{N_y+l}$.
	If either $\hat{\sigma}_y(\mathbf{x}_m) < t ~ \forall \mathbf{x}_m \in \mathcal{Q}_{M_\mathbf{x}-l}$ or the query set is empty, the overall algorithm stops.
	Otherwise, we sample another batch of $l \leq L$ query vectors as described above.

\section{Methodology}
\label{sec:methods}

\subsection{Molecular Reference Systems}

	We considered three different data sets (Table~\ref{tab:data}).
	The first set, \textsf{S13x8}, consists of 13 chemically distinct dimers at 8 different intermolecular distances each (104 dimers in total).
	It is the dispersion-dominated subset (dimers \#34 to \#46) of the \textsf{S66x8} data set,\cite{rezac2011} which can be retrieved from the BEGDB database.\cite{rezac2008}
	The second set, \textsf{ROTA}, contains 1,100 ethyne--pentane dimers with 16 different centroid distances (3.5--10.0~{\AA}).
	The relative orientation between ethyne and pentane was sampled from a uniform spherical distribution (see the Supporting Information for further details).
	All of these dimers contain the same pentane conformer (the one of the ethyne--pentane dimer at equilibrium).
	The third set, \textsf{CONF}, contains 44 ethyne--pentane dimers with a fixed centroid distance of 5.2~{\AA}. 
	Here, random conformations of pentane were sampled with \textsc{RDKit}\cite{landrum2006} and the relative orientation between ethyne and pentane was, again, sampled from a uniform spherical distribution.
	The latter two sets, \textsf{ROTA} and \textsf{CONF}, were created for this work.
	Both molecular structures and interaction energies of all 1,248 dimers considered are provided in the Supporting Information.

\begin{table}
\begin{center}
\caption{
	Overview of the 1,248 molecular reference systems (dimers) studied in this work.
}
\begin{tabular}{lrp{10cm}}
\toprule
Data set & \#Dimers & Description \\
\midrule
\textsf{S13x8} & 104 & dispersion-dominated subset (dimers \#34 to \#46) of the \textsf{S66x8} data set\cite{rezac2011}\\
\textsf{ROTA} & 1,100 & ethyne--pentane dimers; varying relative orientations and centroid distances (3.5--10.0~{\AA}), fixed pentane conformation\\
\textsf{CONF} & 44 & ethyne--pentane dimers; varying relative orientations and pentane conformations, fixed centroid distance (5.2~{\AA})\\
\bottomrule
\end{tabular}
\label{tab:data}
\end{center}
\end{table}

\subsection{Calculation of Interaction Energies}

\subsubsection{Electronic-Structure Calculations}
\label{sec:qc}

	All quantum chemical calculations were carried out with \textsc{Orca} 4.0.1.\cite{neese2012, neese2018} 
	For all chemical systems studied, we performed DLPNO-CCSD(T)\cite{riplinger2013, riplinger2013a} and PBE calculations\cite{perdew1996} on both the interacting and the isolated monomers.
	In the DLPNO-CCDS(T) calculations, we employed the aug-cc-pV$n$Z\cite{dunning1989, kendall1992} basis sets ($n =$ T, Q) and aug-cc-pV$m$Z\cite{neese2003, weigend2006} auxiliary basis sets ($m = n+1$, i.e., $m =$ Q, 5).
	We chose the \texttt{TightPNO} keyword, which is recommended for studying noncovalent interactions.\cite{liakos2015}
	We extrapolated the triple-$\zeta$ and quadruple-$\zeta$ DLPNO-CCSD(T) energies to the complete-basis-set (CBS) limit with the extrapolation scheme developed by Halkier et~al.\cite{halkier1998, halkier1999}~that has already been applied for the calculation of DLPNO-CCSD(T)/CBS energies.\cite{minenkov2017, calbo2017, husch2018}
	The scheme comprises two separate extrapolations for the Hartree--Fock (HF) and correlation (corr) fractions of the electronic energy,
	
\begin{eqnarray}
    \label{eq:cbs}
    E_{\text{elec},\mathcal{S}}^\text{DLPNO-CCSD(T)/CBS} \approx \frac{E_{\text{elec},\mathcal{S}}^\text{HF/QZ} \exp\{1.63X\} - E_{\text{elec},\mathcal{S}}^\text{HF/TZ} \exp\{1.63(X-1)\}}{\exp\{1.63X\} - \exp\{1.63(X-1)\}} \nonumber \\ + \frac{E_{\text{elec},\mathcal{S}}^\text{corr/QZ} X^3 - E_{\text{elec},\mathcal{S}}^\text{corr/TZ} (X-1)^3}{X^3 - (X-1)^3} \ .
\end{eqnarray}

	Here, the subscript $\mathcal{S}$ denotes the molecular system and $X$ is the cardinal number of the larger basis set, i.e., $X = n_\text{max}$ (here, $n_\text{max} = 4$).
	For our PBE calculations, we chose the ma-def-QZVPP\cite{weigend2005, zheng2011} basis set and the def2-QZVP\cite{weigend2006} auxiliary basis set.
	We applied counterpoise (CP) corrections\cite{boys1970} and \texttt{TightSCF} criteria to both DLPNO-CCSD(T) and PBE calculations. 
	The \textit{intermolecular} interaction energy was obtained as the difference between the electronic energies of the isolated monomers $\mathcal{A}$ and $\mathcal{B}$, and the corresponding dimer $\mathcal{AB}$,
	
\begin{equation}
	E_\text{int}^\mathcal{M} \equiv \Delta E_\text{elec}^\mathcal{M} = E_{\text{elec},\mathcal{AB}}^\mathcal{M} - \Big( E_{\text{elec},\mathcal{A}}^\mathcal{M} + E_{\text{elec},\mathcal{B}}^\mathcal{M} \Big) \ ,
\end{equation}
	
	where $\mathcal{M} := \{$DLPNO-CCSD(T)/CBS, PBE/ma-def2-QZVPP$\}$.
	
\subsubsection{D3-Type Dispersion Corrections}

	All D3-type dispersion corrections to PBE interaction energies were obtained from Grimme's stand-alone program \textsc{dftD3}\cite{grimme2010, grimme2011} and incorporate the Becke--Johnson (BJ) damping scheme,\cite{grimme2011}

\begin{equation}
	\label{eq:d3bj}
	E_{\text{disp},\mathcal{S}}^\text{D3(BJ)} \equiv E^\text{D3(BJ)} = \sum_{I > J}  \sum_{n=6,8} s_n \frac{C_n^{IJ}}{R_{IJ}^n + \left(a_1 \sqrt{C_8^{IJ} / C_6^{IJ}} + a_2\right)^n} \ .
\end{equation}

    Here, $R_{IJ}$ is the distance between atoms $I$ and $J$ in system $\mathcal{S}$, and $C_6^{IJ}$ and $C_8^{IJ}$ are the corresponding dispersion coefficients taken from Refs.~\citeonline{grimme2010}~and~\citeonline{grimme2011}, which are negative by definition.
	Spatial position and proton number of the $I$-th atom are denoted $\mathbf{R}_I$ and $Z_I$, respectively.
	The scaling factor $s_6$ is set to unity by definition,\cite{grimme2011} whereas $a_1$, $a_2$, and $s_8$ are functional-dependent, empirical parameters.
	We employed several parametrizations of the D3(BJ) model;\cite{grimme2011, smith2016, weymuth2018} see also Section \ref{sec:methods_d3param}.
	D3(BJ) dispersion corrections were calculated for both interacting and isolated monomers to yield relative D3(BJ) dispersion corrections,
	
\begin{equation}
	\Delta E_\text{disp}^\text{D3(BJ)} = E_{\text{disp},\mathcal{AB}}^\text{D3(BJ)} - \Big( E_{\text{disp},\mathcal{A}}^\text{D3(BJ)} + E_{\text{disp},\mathcal{B}}^\text{D3(BJ)} \Big) \ .
\end{equation}

\subsection{Gaussian Process Regression}

	We employed the \textsc{gpml}\cite{rasmussen2010} code written in \textsc{Matlab}\cite{matlab2018a} for all GP regression and prediction tasks discussed in this work.
	Optimal hyperparameters for the GP models considered were obtained with Rasmussen's implementation of the conjugate gradient algorithm\cite{rasmussen2010} in \textsc{gpml}.
	For estimating the GP posterior distribution, we employed a zero-mean prior and a Gaussian likelihood.

\subsubsection{D3-GP-Type Dispersion Corrections}
\label{sec:methods_d3gp}

	The \textit{residual} (intermolecular) interaction energy,
	
\begin{equation}
\label{eq:residual}
	\Delta E_\text{int} = 
	E_\text{int}^\text{DLPNO-CCSD(T)/CBS} -
	\left(E_\text{int}^\text{PBE/ma-def2-QZVPP} + \Delta E_\text{disp}^\text{D3(BJ)}\right ) \ ,
\end{equation}

    served as the target variable.
    We considered two different featurizations, \textsf{eigD3(BJ)} and \textsf{histD3(BJ)}, both being derived from the D3(BJ) matrix $\mathbf{E}^\text{D3(BJ)}$ with elements

\begin{equation}
\label{eq:d3matrix}
	E_{IJ}^{\text{D3(BJ)}} =
	\begin{cases}
		\sum_{n=6,8} s_n \frac{\displaystyle C_n^{IJ}}{\displaystyle R_{IJ}^n + \left(a_1 \sqrt{C_8^{IJ} / C_6^{IJ}} + a_2\right)^n} & \forall I \neq J \\
		0 & \forall I = J
	\end{cases} \ .
\end{equation}

	To obtain the \textsf{eigD3(BJ)} feature vector, the eigenvalues of $\mathbf{E}^\text{D3(BJ)}$ were sorted according to decreasing absolute value.
	As the dimensionality of the \textsf{eigD3(BJ)} vector represents the number of atoms in the corresponding molecular system, \textsf{eigD3(BJ)} vectors of smaller systems were padded with zeros to match the dimensionality of \textsf{eigD3(BJ)} vectors corresponding to larger systems. 
	This procedure has been adopted from Rupp et~al.~who generated feature vectors from Coulomb matrices.\cite{rupp2012}
	
	By contrast, the \textsf{histD3(BJ)} feature vector is strictly 16-dimensional where each element represents a sum over $E_{IJ}^{\text{D3(BJ)}}$ corresponding to a predefined range of interatomic distances (Table~S1),

\begin{equation}
	e_m^\mathsf{histD3(BJ)} = \sum_{\substack{I > J \\ r_m^\text{min} < R_{IJ} \le r_m^\text{max}}} \sum_{n=6,8} s_n \frac{C_n^{IJ}}{R_{IJ}^n + \left(a_1 \sqrt{C_8^{IJ} / C_6^{IJ}} + a_2\right)^n} \ .
\end{equation}
	
	Both \textsf{eigD3(BJ)} and \textsf{histD3(BJ)} vectors were determined for the interacting and isolated monomers.
	The final feature vectors employed in GP regression read
	
\begin{equation}
	\mathbf{x}^\mathsf{eig/histD3(BJ)} \equiv \Delta \mathbf{e}^\mathsf{eig/histD3(BJ)} = \mathbf{e}_\mathcal{AB}^\mathsf{eig/histD3(BJ)} - \Big( \mathbf{e}_\mathcal{A}^\mathsf{eig/histD3(BJ)} + \mathbf{e}_\mathcal{B}^\mathsf{eig/histD3(BJ)} \Big) \ .
\end{equation}

    We considered both the isotropic Mat\'{e}rn-$\sfrac{1}{2}$ kernel (also known as exponential kernel or Laplacian kernel),\cite{rasmussen2006}
	
\begin{equation}
	k^\text{M12}_\text{iso}(\mathbf{x}_i,\mathbf{x}_j) = \alpha_1 \exp \left\{- d_\text{iso}(\mathbf{x}_i,\mathbf{x}_j) \right\} \ ,
	\label{eq:matern12}
\end{equation}

\begin{equation}
\label{eq:d_iso}
	d_\text{iso}(\mathbf{x}_i,\mathbf{x}_j) = \sqrt{\frac{(\mathbf{x}_i - \mathbf{x}_j) (\mathbf{x}_i - \mathbf{x}_j)^\top}{\alpha_2}} \ ,
\end{equation}

    and the isotropic Mat\'{e}rn-$\sfrac{3}{2}$ kernel,\cite{rasmussen2006}
    
\begin{equation}
	k^\text{M32}_\text{iso}(\mathbf{x}_i,\mathbf{x}_j) = \alpha_1 \left(1 +  \sqrt{3}d_\text{iso}(\mathbf{x}_i,\mathbf{x}_j) \right) \exp \left\{- \sqrt{3}d_\text{iso}(\mathbf{x}_i,\mathbf{x}_j) \right\} \ .
	\label{eq:matern32}
\end{equation}

	As the numerical uncertainty in the interaction energies we consider can be controlled by the SCF convergence criterion, we have prior knowledge of the noise hyperparameter, $\alpha_0$.
	Therefore, we exclude $\alpha_0$ from hyperparameter optimization and set it to a constant value of $1.0 \times 10^{-5}$~kcal~mol$^{-1}$.
	This way, hyperparameter optimization can be accelerated and is less prone to overfitting.
	We optimized the remaining hyperparameters $\alpha_1$ and $\alpha_2$ by minimizing the leave-one-out negative log marginal lihood (maximum of 1,000 optimization steps).
	
\subsubsection{D3(BJ) Reparametrization}
\label{sec:methods_d3param}

    We applied Bayesian optimization\cite{shahriari2016} to reparametrize the D3(BJ) model for PBE.
	We chose the mean absolute relative error (MARE, Eq.~(\ref{eq:mare})), of $\Delta E_\text{int}$ over a subset of \textsf{S13x8} (\textsf{S13x8-T}, see the Supporting Information), as objective function, $y(\mathbf{x})$. 
	Here, $\mathbf{x}$ is a 3-dimensional vector composed of the empirical D3(BJ) parameters $a_1$, $a_2$, and $s_8$.
	25 parameter vectors (initial training set) were generated by Latin hypercube sampling\cite{mckay1979} over the domain $\mathcal{X} = \mathcal{X}_{a_1} \times \mathcal{X}_{a_2} \times \mathcal{X}_{s_8}$, where ${\mathcal{X}_{a_1} := [0.0,0.7]}$, ${\mathcal{X}_{a_2} := [4.0,6.5]}$~{bohr}, and ${\mathcal{X}_{s_8} := [0.0,3.5]}$.
	We rounded each value to the fourth decimal place according to previous D3 parametrizations.
	We evaluated $y(\mathbf{x})$ for each of these 25 parameter triples to approximate the objective function over the same domain by GP regression with the anisotropic Mat\'{e}rn-$\sfrac{3}{2}$ kernel,\cite{rasmussen2006}
	
\begin{equation}
	k^\text{M32}_\text{aniso}(\mathbf{x}_i,\mathbf{x}_j) = \alpha_1 \left(1 +  \sqrt{3}d_\text{aniso}(\mathbf{x}_i,\mathbf{x}_j) \right) \exp \left\{- \sqrt{3}d_\text{aniso}(\mathbf{x}_i,\mathbf{x}_j) \right\} \ ,
	\label{eq:matern32ard}
\end{equation}

\begin{equation}
	d_\text{aniso}(\mathbf{x}_i,\mathbf{x}_j) = \sqrt{(\mathbf{x}_i - \mathbf{x}_j) 
	    \begin{pmatrix} 
		    \alpha_{2,a_1} & 0 & 0 \\ 
		    0 & \alpha_{2,a_2} & 0 \\
		    0 & 0 & \alpha_{2,s_8}
	    \end{pmatrix} 
    (\mathbf{x}_i - \mathbf{x}_j)^\top} \ .
\end{equation}
	
	Here, $\alpha_1$, $\alpha_{2,a_1}$, $\alpha_{2,a_2}$, and $\alpha_{2,s_8}$ represent four hyperparameters in addition to the noise hyperparameter $\alpha_0$ introduced in Eqs.~(\ref{eq:mean})~and~(\ref{eq:var}).
	We optimized the five hyperparameters by minimizing the log marginal likelihood (maximum of 10,000 optimization steps).
	We constructed a lower-confidence-bound acquisition function,
	
\begin{equation}
    u_\text{LCB}(\mathbf{x}) = \hat{\mu}_y(\mathbf{x}) - \hat{\sigma}_y(\mathbf{x}) \ .
\end{equation}

	At each iteration, we defined an equidistant grid with a resolution of 0.1~units over the domain $\mathcal{X}$.
	The initial and terminal grid points of each dimension were located at the bounds of the domain.
	The grid point that minimized $u_\text{LCB}(\mathbf{x})$ defined the center of a cube with an edge length of 0.2~units, being aligned with the three axes of the coordinate system.
	In cases where the cube exceeded the bounds of the domain, we only considered its volume (cuboid) being located inside the domain.
	The faces of this cube or cuboid redefined the bounds of the domain over which we built an equidistant grid with a ten-times-higher resolution of 0.01~units.
	Again, the initial and terminal grid points of each dimension were located at the bounds of the domain.
	We repeated this procedure until we determined the minimum of $u_\text{LCB}(\mathbf{x})$ with a resolution of 0.0001~units.
	Our algorithm stopped after determining the same parameter set twice; here, after 44 iterations.
	The optimal parameter set was used to calculate both D3(BJ) energies and \textsf{eigD3(BJ)}/\textsf{histD3(BJ)} feature vectors.
	
\subsection{Performance Measures} 

    All performance measures considered in this work are defined with respect to the two vectors $\mathbf{y}$ (observations) and $\hat{\mathbf{y}}$ (predictions).
    The mean error (ME) represents the centroid of the error vector $\boldsymbol{\Delta} = \mathbf{y} - \hat{\mathbf{y}}$,
    
\begin{equation}
    \label{eq:me}
    \hat{\mu} \equiv \text{ME} = \frac{1}{N}\sum_{n=1}^N \Delta_n \ .
\end{equation}

    The root-mean-square deviation (RMSD) is a measure of the scatter of $\mathbf{\Delta}$,
    
\begin{equation}
    \hat{\sigma} \equiv \text{RMSD} = \sqrt{\frac{1}{N-1}\sum_{n=1}^N \left(\Delta_n - \hat{\mu}\right)^2} \ .
\end{equation}

    The root-mean-square error (RMSE) can be expressed as a function of both ME and RMSD,
    
\begin{equation}
    \text{RMSE} = \sqrt{\frac{1}{N-1}\sum_{n=1}^N \Delta_n^2} = \sqrt{\frac{N}{N-1}\hat{\mu}^2 + \hat{\sigma}^2} \ .
\end{equation}

    The mean absolute error (MAE), 
    
\begin{equation}
    \label{eq:mae}
    \text{MAE} = \frac{1}{N}\sum_{n=1}^N \left|\Delta_n \right| \ ,
\end{equation}

    is the $L_1$-norm analog to the RMSE.
    All performance measures presented can be transformed into dimensionless quantities by dividing each element of the error vector by the absolute value of the corresponding element in the observation vector,

\begin{equation}
    \delta_n = \frac{\Delta_n}{\left|y_n\right|} \ .
\end{equation}

    All $\{\Delta_n\}$ in Eqs.~(\ref{eq:me})--(\ref{eq:mae}) need to be replaced by $\{\delta_n\}$ to obtain relative-performance measures.
    Examples are the mean relative error,
    
\begin{equation}
    \text{MRE} = \frac{1}{N}\sum_{n=1}^N \delta_n \ ,
\end{equation}

	and the mean absolute relative error,
	
\begin{equation}
    \label{eq:mare}
    \text{MARE} = \frac{1}{N}\sum_{n=1}^N \left|\delta_n \right| \ .
\end{equation}	
	
    Note that relative-performance measures yield unreasonably large values in cases where the absolute value of an observation is smaller  than or comparable to the underlying measurement/numerical uncertainty, i.e., when it is close to zero.
    In such cases, a capped version of the relative error,
    
\begin{equation}
\label{eq:cap}
    \delta_n^c = \frac{\Delta_n}{\max \left\{\left|y_n\right|, c\right\}} \ ,
\end{equation}

	avoids running into singularity issues.
	The cap $c$ is a strictly positive real number.

\section{Results and Discussion}
\label{sec:results}

\subsection{Assessment of Interaction Energies}

    Smith et~al.\cite{smith2016}~included the \textsf{S66x8} data set developed by \v{R}ez\'a\v{c} et~al.\cite{rezac2011}~among other data sets for their reparametrization of PBE-D3(BJ) (other density functional were also considered).
    They also adopted the corresponding CP-corrected CCSD(T)/CBS$^\ast$ interaction energies. 
    We introduce the asterik to distinguish the extrapolation scheme employed by \v{R}ez\'a\v{c} et~al.~from the one considered in this work.
    While Smith et~al.~employed the same basis set (def2-QZVP) as Grimme and co-workers\cite{grimme2011} for calculating PBE interaction energies, they applied CP corrections instead of neglecting basis set superposition errors.
    Considering the \textsf{S13x8} subset, we compared the results presented in this work against those published by Smith et~al.
    In the following, we report all relative and signed errors with respect to their study.
    
    We find a squared correlation coefficient of $r^2 = 0.9999$ between PBE/def2-QZVP (Smith et~al.\cite{smith2016}) and PBE/ma-def2-QZVPP (this work) interaction energies.
    The MAE and the corresponding RMSD both amount to 0.0087~kcal~mol$^{-1}$.
    Given that the CP-corrected CCSD(T)/CBS$^\ast$ interaction energies are reported with an accuracy of 0.01~kcal~mol$^{-1}$, we conclude that the effect of basis set alteration on PBE interaction energies is insignificant in this case.
    
    Furthermore, we find a squared correlation coefficient of $r^2 = 0.9954$ between CCSD(T)/CBS$^\ast$ (\v{R}ez\'a\v{c} et~al.\cite{rezac2011}) and DLPNO-CCSD(T)/CBS (this work) interaction energies.
    MAE and RMSD amount to 0.11~kcal~mol$^{-1}$ and 0.12~kcal~mol$^{-1}$, respectively.
    We consider these deviations significant given the limiting resolution of 0.01~kcal~mol$^{-1}$ mentioned above.
    For the five smallest relative distances (0.90, 0.95, 1.00, 1.05, and 1.10 with respect to the equilibrium distance, $r_\text{eq}$), we find that the unsigned ME ($|$ME$|$) equals the MAE.
    As the ME is negative in all of these cases, the corresponding DLPNO-CCSD(T)/CBS dispersion interactions are, without exception, determined to be less attractive than the ones obtained from CCSD(T)/CBS$^\ast$ calculations.
    For the three largest relative distances (1.25$r_\text{eq}$, 1.50$r_\text{eq}$, and 2.00$r_\text{eq}$), the unsigned ME represents at least 70~\% of the MAE.
    In these cases, the ME is positive, suggesting that DLPNO-CCSD(T)/CBS tends to predict more attractive dispersion corrections at long distances compared to CCSD(T)/CBS$^\ast$.
    However, the ME over the entire \textsf{S13x8} set amounts to $-$0.09~kcal~mol$^{-1}$.

	The CBS extrapolation scheme employed by \v{R}ez\'a\v{c} et~al.\cite{rezac2011}~involves MP2 energies obtained with aug-cc-pVTZ and aug-cc-pVQZ basis sets, and approximates the CBS limit of the correlation energy by adding to the MP2/CBS energy a CCSD(T)/aug-cc-pVDZ correction.
	In this work, however, we extrapolate directly from DLPNO-CCSD(T)/aug-cc-pV\sfrac{T}{Q}Z energies as defined in Eq.~(\ref{eq:cbs}).
	Therefore, it is not only the underlying electronic-structure approximation but also the extrapolation scheme that is different from the methodology reported by \v{R}ez\'a\v{c} et~al.
	While CCSD(T) should by construction capture more long-range interactions than DLPNO-CCSD(T), our results are based on triple/quadruple-$\zeta$ basis sets (aug-cc-pV\sfrac{T}{Q}Z) compared to a double-$\zeta$ basis set (aug-cc-pVDZ) employed by \v{R}ez\'a\v{c} et~al.
	It is debatable whether diffuse basis functions can compensate for the inaccuracies introduced by a double-$\zeta$ basis set.
	The positive ME determined for the three largest intermolecular distances indicates that DLPNO-CCSD(T)/aug-cc-pV\sfrac{T}{Q}Z calculations may capture more long-range interactions than CCSD(T)/aug-cc-pVDZ corrections to MP2/aug-cc-pV\sfrac{T}{Q}Z energies.
	
	To gain more insight, we examined the effect of the CBS extrapolation scheme introduced in Eq.~(\ref{eq:cbs}) on predicting interaction energies (as measured by the MAE over \textsf{S13x8}) by varying its two parameters.
	We find that the interaction energies are insensitive with respect to smaller (more negative) values of the Hartree--Fock parameter (default: $-$1.63, dimensionless, cf.~Eq.~(\ref{eq:cbs})).
	For larger (less negative or positive) values, the MAE increases.
	Increasing the correlation parameter from its default value of $-$3.00 (dimensionless, cf.~Eq.~(\ref{eq:cbs})) to $-$1.17 minimizes the MAE to 0.03~kcal~mol$^{-1}$ (RMSD~$=$ 0.04~kcal~mol$^{-1}$), which constitutes a significant reduction of the deviation between DLPNO-CCSD(T)/CBS and CCSD(T)/CBS$^\ast$ interaction energies.
	Moreover, the overall ME is now slightly positive (0.01~kcal~mol$^{-1}$).
	For comparative purposes, we note that ME, MAE, and RMSD of DLNPO-CCSD(T)/aug-cc-pVQZ interaction energies (with respect to CCSD(T)/CBS$^\ast$) are larger in magnitude than those of their DLNPO-CCSD(T)/CBS analogs.
	These quantities amount to $-$0.13~kcal~mol$^{-1}$, 0.15~kcal~mol$^{-1}$, and 0.16~kcal~mol$^{-1}$, respectively.
	We do not adopt the tuned correlation parameter ($-$1.17) for this study as a reduced deviation from the CCSD(T)/CBS$^\ast$ results over \textsf{S13x8} does not necessarily imply an improvement in accuracy.
	Therefore, we consider a reparametrization of the D3(BJ) model for PBE as it has previously been trained on CCSD(T)/CBS$^\ast$ interaction energies.\cite{grimme2011, smith2016, weymuth2018}

\subsection{System-focused D3(BJ) Reparametrization}
\label{sec:reparametrization}

    A reparametrization of the D3(BJ) model for PBE is not only important for an unbiased assessment of its performance with respect to DLPNO-CCSD(T)/CBS reference interaction energies.
    As the \textsf{eigD3(BJ)} and \textsf{histD3(BJ)} feature vectors (cf.~Section \ref{sec:methods_d3gp}) are functions of the empirical D3(BJ) parameters $a_1$, $a_2$, and $s_8$, mean and variance obtained from GP regression will also depend on the specific D3(BJ) parameters.
    A system-focused reparametrization would, therefore, ensure an unbiased performance test of the D3(BJ)-GP model as we expect to obtain close-to-optimal PBE-D3(BJ) interaction energies and, hence, maximally reduced residual interaction energies, Eq.~(\ref{eq:residual}), as input for GP regression.
    Following the line of argument by Smith et~al.,\cite{smith2016} we decided to optimize the D3(BJ) parameters for PBE with respect to a relative rather than an absolute error, as the latter tends to bias the parameters toward systems with large absolute interaction energies.
    In this work, for instance, the largest unsigned DLPNO-CCSD(T)/CBS interaction energy is 320 times larger than the smallest one.
    
\begin{figure}
    \includegraphics[width=\textwidth]{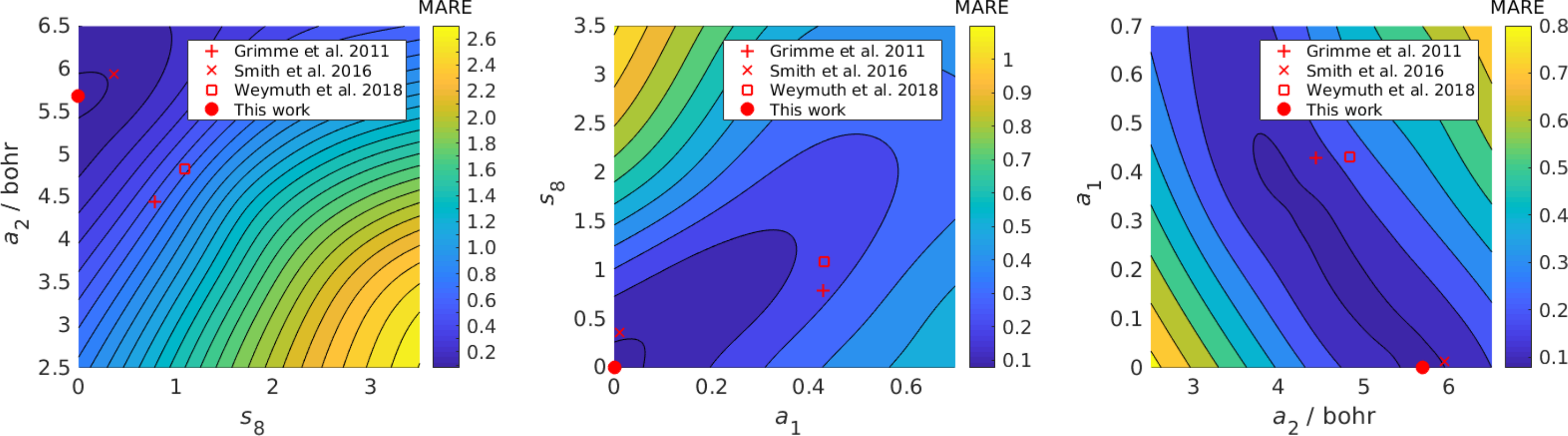}
    \caption{
    MARE for PBE-D3(BJ) interaction energies as a function of the three empirical D3(BJ) parameters $a_1$, $a_2$, and $s_8$.
    The bounds of the contour plots represent the bounds specified for the optimization procedure.
    The red dots represent the global minimum of the MARE hypersurface, which was determined with respect to the \textsf{S13x8-T} set.
    Three additional parameter triples are displayed for comparison.
    To produce this figure, $a_1$ and $s_8$ were set to zero in the left and right panels, respectively; in the central panel, $a_2$ took a value of 5.6841 bohr.
    }
    \label{fig:contour}
\end{figure}

    The objective function (MARE) of the optimization procedure is shown in Figure~\ref{fig:contour}.
    At the global minimum, $a_1$ and $s_8$ are zero, and $a_2 = 5.6841$~bohr.
    This remarkable reduction of the number of parameters may be explained with the underlying training set (\textsf{S13x8-T}), which solely consists of dispersion-dominated dimers.
    The training sets employed in previous D3(BJ) parametrizations\cite{grimme2011, smith2016, weymuth2018} were more heterogeneous and also included, e.g., hydrogen-bonded aggregates.
    The smaller the number of effects that need to be captured, the more likely it is that a simple model captures these effects as accurately as a more complex model.
    To support this hypothesis, we repeated the optimization with the identical settings and training systems, but this time, we incorporated the CCSD(T)/CBS$^\ast$ interaction energies published by \v{R}ez\'a\v{c} et~al.\cite{rezac2011}
    Again, $a_1$ and $s_8$ are zero at the global minimum, and $a_2$ changes to a slightly lower value of 5.5530~bohr. We do not consider this parameter set in the following.
    These findings suggest that a simplified expression of the D3(BJ) dispersion energy, defined in Eq.~(\ref{eq:d3bj}), can be applied when studying dispersion-dominated systems (at least in the case of PBE),
    
\begin{equation}
	E^\text{D3(BJ)} = \sum_{I > J} \frac{C_6^{IJ}}{R_{IJ}^6 + a_2^6} \ .
\end{equation}

    This expression does no longer involve dipole--quadrupole interactions ($n = 8$), neither through the dipole--quadrupole term itself nor through the term $\sqrt{C_8^{IJ} / C_6^{IJ}}$ of the damping function, which is multiplied by $a_1$.
    For $R_{IJ} \rightarrow 0$, the corresponding D3(BJ) term converges to the finite negative value of the simple expression $C_6^{IJ} / a_2^6$.
    
    As the D3(BJ) parameter set for PBE is a function of the training set (here, \textsf{S13x8-T}), the reference interaction energies (here, $E_\text{int}^\text{DLPNO-CCSD(T)/CBS}$), and the objective (here, MARE), it is not surprising that optimal parameter sets obtained in other studies may differ significantly from the one determined in this study.
    For instance, Grimme et~al.\cite{grimme2011}~and our group\cite{weymuth2018} chose a weighted MAE as objective, CCSD(T)/CBS$^\ast$ energies, and a training set comprising equilibrium structures only.
    Whereas Smith et~al.\cite{smith2016}~also incorporated CCSD(T)/CBS$^\ast$ energies, they employed a similar objective (a capped version of the MARE, cf.~Eq.~(\ref{eq:cap})) and a training set that also included nonequilibrium structures.
    These methodological differences may explain why the PBE-specific D3(BJ) parameters by Smith et~al.~are much more similar to the ones obtained in this work compared to the parameters published by Grimme et~al.\cite{grimme2011}~and previously by our group\cite{weymuth2018} (note that these parameter sets are, in turn, similar to each other).
    The similarity to the parameters by Smith et~al.~underpins once more that the change from CCSD(T)/CBS$^\ast$ to DLPNO-CCSD(T)/CBS has only a minor effect on the D3(BJ) model.
    
    The accuracy of the D3(BJ) parameter sets mentioned above with respect to three different validation sets (\textsf{S13x8-V}, \textsf{ROTA}, and \textsf{CONF}) are reported in Table~\ref{tab:d3param}.
    The validation sets considered have no data points in common with the training set (\textsf{S13x8-T}).
    Our new D3(BJ) parameters for PBE result in the lowest MAREs and MAEs over all validation sets (in the case of \textsf{S13x8-V}, the parameters by Smith et~al.~yield the same MAE).
    The new parameter set also leads to the smallest unsigned MRE ($|$MRE$|$) independent of the validation set and, hence, reveals the strongest reduction of systematic errors in D3(BJ)-type dispersion corrections.
    Based on this problem-specific validation, we conclude that our new PBE-specific D3(BJ) parameter set maximizes the statistical validity of D3(BJ)-type dispersion corrections for the given domain of application.
    If not otherwise mentioned, the term ``D3(BJ)'' and all other terms including it refer to that parameter set, i.e., $a_1 = s_8 = 0$ and $a_2 = 5.6841$~bohr.
    
\begin{table}[ht!]
    \begin{center}
    \small
    \caption{
    Performances (as measured by MARE and MAE) of four different PBE-specific D3(BJ) parameter sets with respect to three validation sets (\textsf{S13x8-V}, \textsf{ROTA}, and \textsf{CONF}).
    The highest accuracies are highlighted in bold.
    The D3(BJ) parameters $a_1$ and $s_8$ are dimensionless. 
    Values for the D3(BJ) parameter $a_2$ and the MAE are reported in bohr and kcal~mol$^{-1}$, respectively.
    }
    \begin{tabular}{lrrrrrrrrrrrr}
    \toprule
    & & & && \multicolumn{2}{c}{\textsf{S13x8-V}} && \multicolumn{2}{c}{\textsf{ROTA}} && \multicolumn{2}{c}{\textsf{CONF}} \\\cmidrule{6-7}\cmidrule{9-10}\cmidrule{12-13}
    Reference & $a_1$ & $a_2$ & $s_8$ && MARE & MAE && MARE & MAE && MARE & MAE\\
    \midrule
    Grimme et~al.\cite{grimme2011} & 0.4289 & 4.4407 & 0.7875 && 13.4~\% & 0.27 && 27.8~\% & 0.10 && 34.7~\% & 0.21 \\
    Smith et~al.\cite{smith2016} & 0.0121 & 5.9390 & 0.3589 && 9.5~\% & \textbf{0.19} && 23.8~\% & 0.08 && 29.7~\% & 0.18 \\
    Weymuth et~al.\cite{weymuth2018} & 0.4309 & 4.8327 & 1.0892 && 12.0~\% & 0.26 && 25.6~\% & 0.09 && 30.8~\% & 0.19 \\
    This work & 0.0000 & 5.6841 & 0.0000 && \textbf{8.2~\%} & \textbf{0.19} && \textbf{20.1~\%} & \textbf{0.07} && \textbf{23.8~\%} & \textbf{0.15} \\
    \bottomrule
    \end{tabular}
    \label{tab:d3param}
    \end{center}
\end{table}

\subsection{Learning Curves of the D3(BJ)-GP Model}
\label{sec:learning-curves}
 
\begin{figure}
    \centering
    \begin{minipage}{.5\textwidth}
        \centering
        \includegraphics[width=0.9\textwidth]{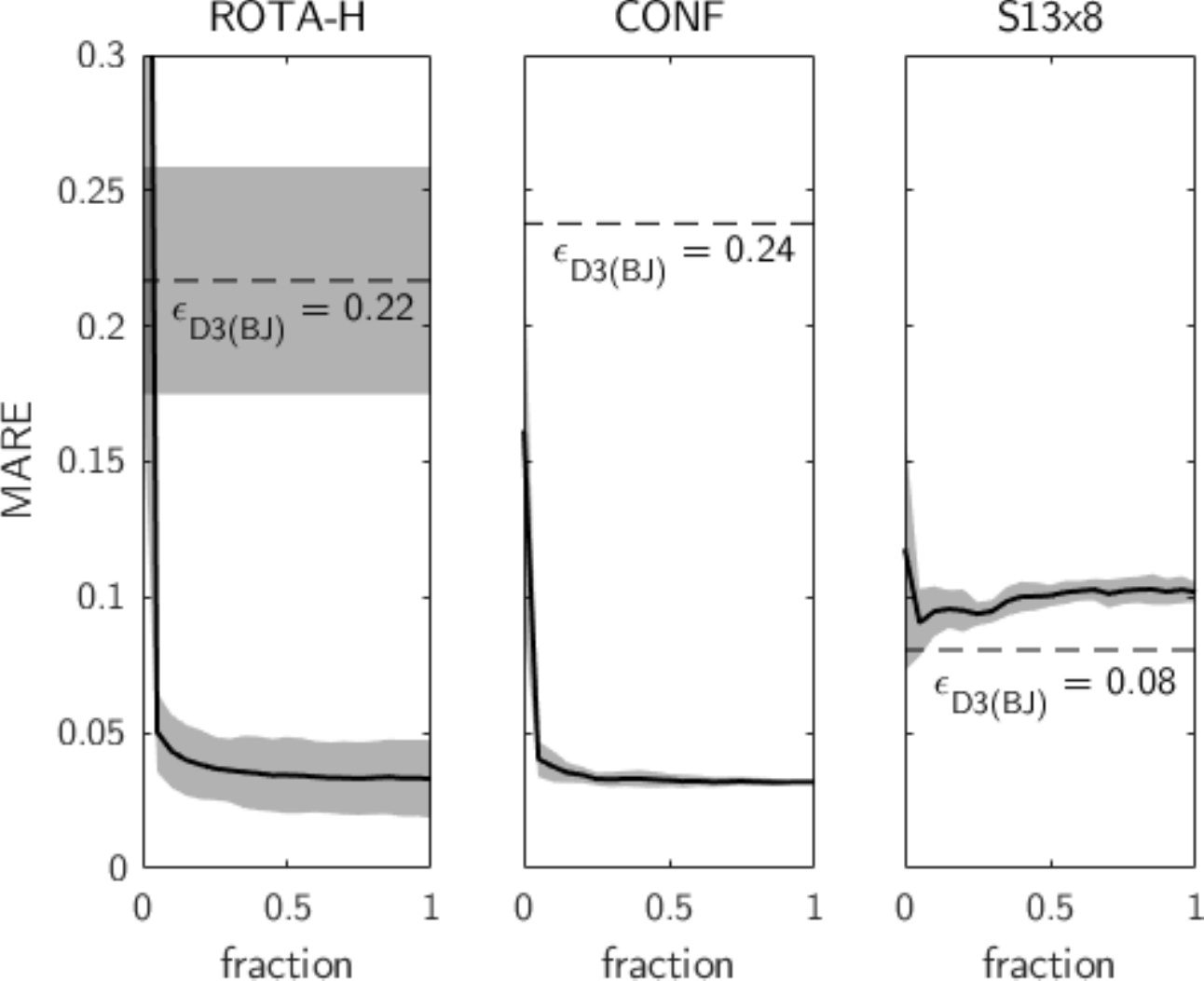}
        \caption*{\textbf{a)} Mat\'{e}rn-$\sfrac{1}{2}$}
    \end{minipage}%
    \begin{minipage}{0.5\textwidth}
        \centering
        \includegraphics[width=0.9\textwidth]{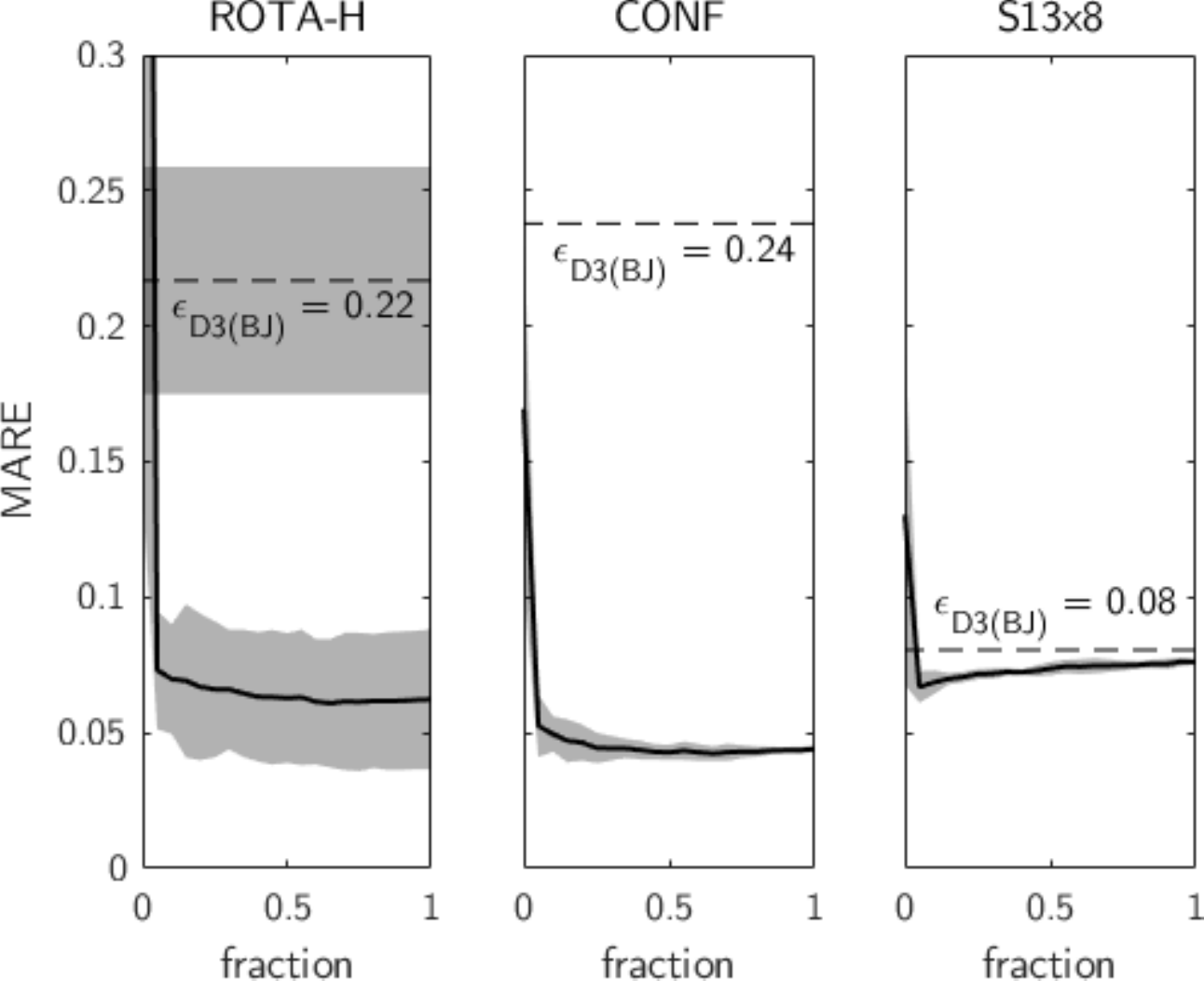}
        \caption*{\textbf{b)} Mat\'{e}rn-$\sfrac{3}{2}$}
    \end{minipage}
    \caption{
    Learning curves (solid lines) of the D3(BJ)-GP model for PBE with respect to three disjoint test sets (\mbox{\textsf{ROTA-H}}, \textsf{CONF}, and \textsf{S13x8}).
    The learning curves (as measured by the MARE) refer to \textsf{histD3(BJ)} features and two kernels, 
    \textbf{a)} Matérn-\sfrac{1}{2} and \textbf{b)} Matérn-\sfrac{3}{2}.
    The D3(BJ)-GP model was trained on the \textsf{ROTA-T} set, a random sample of 1,000 instances drawn from the \textsf{ROTA} set (the remaining 100 instances are contained in the holdout set \textsf{ROTA-H}).
    The training set fractions were taken in 20 equidistant steps between 1~\% and 100~\%.
    Furthermore, the entire training set (1,000 data points) was drawn 10 times, resulting in standard deviations indicated by the gray bands.
    The errors of the corresponding D3(BJ)-corrected PBE interaction energies, denoted $\epsilon_\text{D3(BJ)}$, are shown as horizontal dashed lines.
    }
    \label{fig:learning-curves}
\end{figure}

    To examine GP regression for interpolation and extrapolation tasks, we present three prediction scenarios with the same underlying training set on three different test sets (Figure~\ref{fig:learning-curves}).
    Recall that the target variable is the residual interaction energy, $\Delta E_\text{int}$, defined in Eq.~(\ref{eq:residual}).
    For the training set, 1,000 dimers were randomly drawn (10 repetitions) from \textsf{ROTA}, which we refer to as \textsf{ROTA-T}.
    Its complementary holdout set consisting of the 100 remaining \textsf{ROTA} dimers, denoted \textsf{ROTA-H}, is used as the first test set.
    The other two test sets are \textsf{CONF} and \textsf{S13x8}.
    Starting from a 1~\% fraction of \textsf{ROTA-T} data, we successively added data points in 20 equidistant intervals until the entire \textsf{ROTA-T} set was used for training.
    After each addition, we performed a hyperparameter optimization.
    We determined the MARE of both PBE-D3(BJ) interaction energies (independent of the training set and, hence, constant) and the corresponding PBE-D3(BJ)-GP interaction energies.
    For the latter, we employed the \textsf{histD3(BJ)} featurization.
    Furthermore, we considered two different kernels: The Mat\'{e}rn-\sfrac{1}{2} (Laplacian/exponential) kernel (Figure~\ref{fig:learning-curves}a) and the Mat\'{e}rn-\sfrac{3}{2} kernel (Figure~\ref{fig:learning-curves}b) introduced in Eqs.~(\ref{eq:matern12})~and~(\ref{eq:matern32}), respectively.
    During preliminary testing, we found that these kernels perform best across the Mat\'{e}rn kernel family (to which also the Gaussian/squared exponential kernel belongs\cite{rasmussen2006}) for predicting residual interaction energies.
    
    We find that the D3(BJ)-GP model for PBE (solid lines) almost immediately\,---\,after training on less than five data points\,---\,predicts improved dispersion corrections compared to D3(BJ) (dashed lines) for both the \textsf{ROTA-H} and \textsf{CONF} test sets.
    The predictions improve at a high rate until about 30 data points of the training set (\textsf{ROTA-T}) are considered.
    Afterwards, the predictions continue to improve only marginally.
    The small associated standard deviations (gray bands) indicate that the learning curves are rather insensitive toward the actual composition of the training set.
    While the Matérn-\sfrac{1}{2} kernel leads to slightly smaller prediction errors (3~\%) and standard deviations ($\leq$1~\%) than the Matérn-\sfrac{3}{2} kernel (5--6~\% and $\leq$2~\%, respectively), both kernels yield qualitatively identical results.
    Note that, since \textsf{ROTA-H} is complementary to \textsf{ROTA-T}, the PBE-D3(BJ) error for \textsf{ROTA-H} is also associated with a nonzero standard deviation.
    While each dimer included in the training set (\textsf{ROTA-T}) contains the same pentane conformer, the GP model can, without any delay, generalize to 44 other pentane conformers contained in the \textsf{CONF} dimers.
    This finding suggests that the \textsf{CONF} set has a large overlap with the \textsf{ROTA} set in feature space (cf.~Section~\ref{sec:feature_space}).

    Regarding the \textsf{S13x8} test set, the D3(BJ)-GP model based on the Matérn-\sfrac{1}{2} kernel results in consistently larger prediction errors than the D3(BJ) model.
    By contrast, the D3(BJ)-GP model based on the Matérn-\sfrac{3}{2} kernel quickly improves over the D3(BJ) model, even though only slightly.
    Beyond a 5~\% fraction of training data, adding more data to \textsf{ROTA-T} does not improve predictions in both cases.
    Given that the \textsf{S13x8} set contains chemically different systems compared to \textsf{ROTA} and \textsf{CONF} (who are solely composed of ethyne--pentane dimers), we consider this test an extrapolation task.
    To understand this result, we analyzed the hyperparameter $\alpha_2$ introduced in Eq.~(\ref{eq:d_iso}) for both kernels, which describes on what length scale in feature space information decays.
    The smaller the length scale, the higher is the decay rate of the kernel $k(\mathbf{x}_i,\mathbf{x}_j)$ when the distance between $\mathbf{x}_i$ and $\mathbf{x}_j$ increases.
    As we specified a zero-mean prior, the mean of the posterior distribution converges to zero for large distances between $\mathbf{x}_i$ and $\mathbf{x}_j$.
    In such cases, D3(BJ)-GP- and D3(BJ)-type dispersion corrections become identical.
    This property of the posterior distribution is highly appealing as it preserves the accuracy of PBE-D3(BJ) interaction energies when the distance (in feature space) to the training set increases.
    
    For the Matérn-\sfrac{3}{2} kernel, $\alpha_2$ is on the order of 0.1~kcal~mol$^{-1}$.
    Given that mean (and standard deviation) of the \textsf{ROTA} and \textsf{S13x8} reference interaction energies are 0.28~kcal~mol$^{-1}$ (0.25~kcal~mol$^{-1}$) and 2.03~kcal~mol$^{-1}$ (1.19~kcal~mol$^{-1}$), respectively, it seems that most of the \textsf{S13x8} systems have a distance to the \textsf{ROTA} set that is significantly larger than $\alpha_2$.
    To test this hypothesis, we examined the kernel matrix $\mathbf{K}\left(\mathbf{X}^\textsf{ROTA},\mathbf{X}^\textsf{S13x8}\right)$, where $\mathbf{X}^\textsf{ROTA}$ and $\mathbf{X}^\textsf{S13x8}$ represent the feature vectors of training and test sets, respectively.
    The kernel matrix was normalized such that $k(\mathbf{x}_i,\mathbf{x}_i) = 1$ (arbitrary units), which constitutes the maximum possible value of $k(\mathbf{x}_i,\mathbf{x}_j)$.
    Of the 104 dimers contained in the \textsf{S13x8} set, only 24 exhibit a kernel for which $k\left(\mathbf{x}_i^\textsf{ROTA},\mathbf{x}_{j}^\textsf{S13x8}\right) > 0.1$, which indicates that the \textsf{ROTA} and \textsf{S13x8} sets have only a small overlap in feature space (confirming the extrapolation assumption stated above).
    Moreover, for 50~\% of the \textsf{S13x8} dimers, we find that $k\left(\mathbf{x}_i^\textsf{ROTA},\mathbf{x}_{j}^\textsf{S13x8}\right) < 0.0005$.
    Hence, the slight improvement of PBE-D3(BJ)-GP over PBE-D3(BJ) (Figure~\ref{fig:learning-curves}b, right panel) originates from the few \textsf{S13x8} dimers that are close to \textsf{ROTA} in the \textsf{histD3(BJ)} feature space. 
    For the remaining majority of the \textsf{S13x8} test set, the D3(BJ)-GP model yields essentially identical dispersion corrections as its D3(BJ) analog.
    
    By contrast, $\alpha_2$ is on the order of 10$^4$~kcal~mol$^{-1}$ in the Matérn-\sfrac{1}{2} case.
    This length scale is by several orders of magnitude larger than the range of reference interaction energies considered in this work.
    The largest element of the normalized kernel matrix $\mathbf{K}\left(\mathbf{X}^\textsf{ROTA},\mathbf{X}^\textsf{S13x8}\right)$ amounts to 0.9997.
    Hence, the D3(BJ)-GP model can be considered an almost global model just as its D3(BJ) analog.
    That the former performs worse than the latter (Figure~\ref{fig:learning-curves}a, right panel) can be attributed to this extensive length scale and the remaining kernel-specific hyperparameter $\alpha_1$.
    They were learned from samples of the \textsf{ROTA} set, which spans a much smaller energy range than the \textsf{S13x8} set.
    While these hyperparameters perform well for interpolation tasks (cf.~Figure~\ref{fig:learning-curves}a, left and central panels), they perform poorly for extrapolation tasks.
    Note that the deviation of the \textsf{S13x8}-related learning curve to the PBE-D3(BJ) error (dashed line) is significantly larger than the standard deviation of the learning curve (gray band).
    This behavior is an indication of underfitting; the prediction model is not yet flexible enough to generalize to data outside the training domain.
    
    The results discussed above and shown in Figure~\ref{fig:learning-curves} were derived from the \textsf{histD3(BJ)} featurization.
    The complementary \textsf{eigD3(BJ)} results are reported in the Supporting Information (Figure~S1).
    They are in qualitative agreement with the \textsf{histD3(BJ)} results, but mean and standard deviation of the \textsf{eigD3(BJ)} learning curves are systematically larger than those of the \textsf{histD3(BJ)} learning curves.
    This relatively poor performance can be explained in part with the properties of the D3(BJ) matrix, introduced in Eq.~(\ref{eq:d3matrix}), of the interacting system.
    As its diagonal-block elements are effectively zero, each nonzero element in the \textsf{eigD3(BJ)} vector with an odd index is nearly identical in magnitude to the next (evenly indexed) element.
    They only differ by their signs, i.e., there is always a pair of one positive eigenvalue and one negative eigenvalue which share almost identical absolute values.
    The order of signs seems to be randomly flipped, which introduces some degree of noise to the \textsf{eigD3(BJ)} vector.
    Applying a strict order of signs led to systematically smaller learning curves compared to a random order, but they are still systematically larger than the \textsf{histD3(BJ)} learning curves.
    
    We also analyzed the learning curves of a linear model with respect to the same training and test sets (Figure~S2).
    Both \textsf{histD3(BJ)} and \textsf{eigD3(BJ)} feature vectors served (separately) as input variables.
    In all cases, the learning curves obtained with the linear models exhibit significantly larger mean values and standard deviations than those obtained with our GP models.
    We also explored other polynomial models (constant, quadratic, cubic, and quartic); the results are of the same quality.
    Please see the Supporting Information for further details.
    We conclude that linear regression based on polynomial models is not well suited for the prediction tasks considered in this study.
    
\subsection{Data Diversity Analysis}
\label{sec:feature_space}

\begin{figure}
    \centering
    \includegraphics[width=\textwidth]{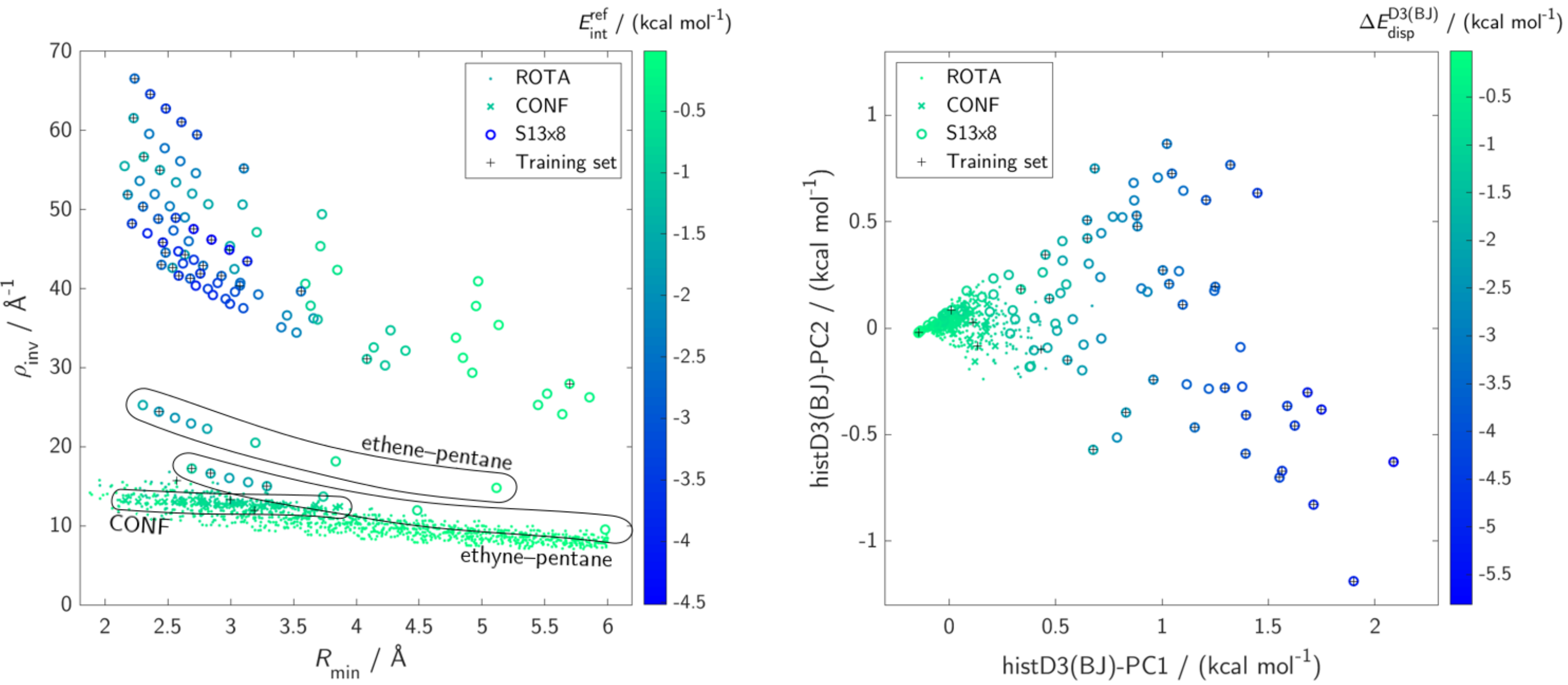}
    \caption{
    Two different representations of the data sets considered in this study: \textsf{ROTA} (dots), \textsf{CONF} (diagonal crosses), and \textsf{S13x8} (circles). 
    Left panel: The reference interaction energy $E_\text{int}^\text{ref}$ (ref~$=$ DLPNO-CCSD(T)/CBS) is shown as a function of two intermolecular-distance measures: the minimum interatomic distance, $R_{\mathrm{min}}$, and the cumulative reciprocal interatomic distance, $\rho_\text{inv}$, of Eq.~(\ref{eq:rho-inv}).
    From this overall data set comprising 1,248 dimers, 40 dimers were selected by BVS and are given as black crosses.
    Right panel: The intermolecular dispersion energy $\Delta E_\text{disp}^\text{D3(BJ)}$ is shown as a function of the first two principal components (\textsf{-PC1} and \textsf{-PC2}) derived from the set of \textsf{histD3(BJ)} feature vectors.
    The first principal component correlates almost perfectly with $\Delta E_\text{disp}^\text{D3(BJ)}$ ($r^2 = 0.9917$).
    The training sets (black crosses) shown in both panels are identical.
    }
    \label{fig:scatter}
\end{figure}

    To interpret the results of the previous section from a data perspective, we put the different data sets (\textsf{ROTA}, \textsf{CONF}, and \textsf{S13x8}) into a mutual context by analyzing the distributions of their molecular systems over different variables.
    As the intermolecular distance and the relative spatial orientation both have an essential effect on the magnitude of noncovalent interactions, an intuitive way to structure the data is to unravel these two properties (Figure~\ref{fig:scatter}a).
    For this purpose, we introduce two measures of the intermolecular distance: the minimum interatomic distance, $R_\text{min}$, and the cumulative reciprocal interatomic distance,

\begin{equation}
    \rho_\text{inv} = \sum_{I \in \mathcal{A}} \sum_{J \in \mathcal{B}} R_{IJ}^{-1} \ .
    \label{eq:rho-inv}
\end{equation}
    
    While $R_\text{min}$ correlates with the largest interatomic noncovalent interaction between two molecules, $\rho_\text{inv}$ encodes the overall noncovalent interactions between two molecules. 
    This data decomposition confirms that (i) \textsf{CONF} has a large overlap with \textsf{ROTA} and (ii) \textsf{S13x8} is well-separated from \textsf{ROTA}.
    The uniform sampling of the \textsf{ROTA} dimers with respect to both distance and relative orientation is reflected by its uniformly distributed data cloud.
    The relatively large $\rho_\text{inv}$ values of the \textsf{S13x8} set can be mainly attributed to larger monomer sizes compared to the \textsf{ROTA} and \textsf{CONF} set.
    It is not surprising that the ethyne--pentane and (subsequently) ethene--pentane series of the \textsf{S13x8} set are closest to the \textsf{ROTA}--\textsf{CONF} data cloud in this representation.
    Note that the ethyne--pentane series represents an upper bound of $\rho_\text{inv}$ to the \textsf{ROTA}--\textsf{CONF} cloud, which originates from the parallel orientation of its dimers.
    
    The majority of systems over all data sets lies in the high-energy regime as measured by the intermolecular D3(BJ) dispersion energy.
    Hence, the majority of systems is located in a dense cluster in the \textsf{eigD3(BJ)} and \textsf{histD3(BJ)} feature spaces, respectively.
    The most informative linear representation of a feature space in two dimensions can be obtained by selecting its first two principal components.
    Here, we applied a principal component analysis\cite{hastie2009} to the \textsf{histD3(BJ)} feature space (Figure~\ref{fig:scatter}b).
    Its first principle component reveals an almost perfect correlation with the intermolecular D3(BJ) dispersion energy (${r^2 = 0.9917}$).
    To minimize the number of training data (and hence, the number of electronic-structure calculations) needed to achieve a predefined accuracy over the entire feature domain considered, we would choose a sampling strategy that samples this domain uniformly.
    The black crosses in Figure~\ref{fig:scatter} represent such a training set, which could be obtained from, e.g., $k$-means clustering.\cite{hastie2009}
    By contrast, random sampling would mainly select data located in the \textsf{ROTA}--\textsf{CONF} cluster due to its large number of instances, which would lead to poor generalizations into the low-energy feature subspace.
    However, sampling the feature space uniformly is not sufficient as we would not know how many data points to pick in order to achieve a predefined accuracy.
    For this purpose, we harness the variance of the GP posterior distribution.

\subsection{Variance-Based Selection of Training Data}
\label{sec:bayesian-selection}

\begin{figure} [!ht]
    \centering
    \includegraphics[width=0.7\textwidth]{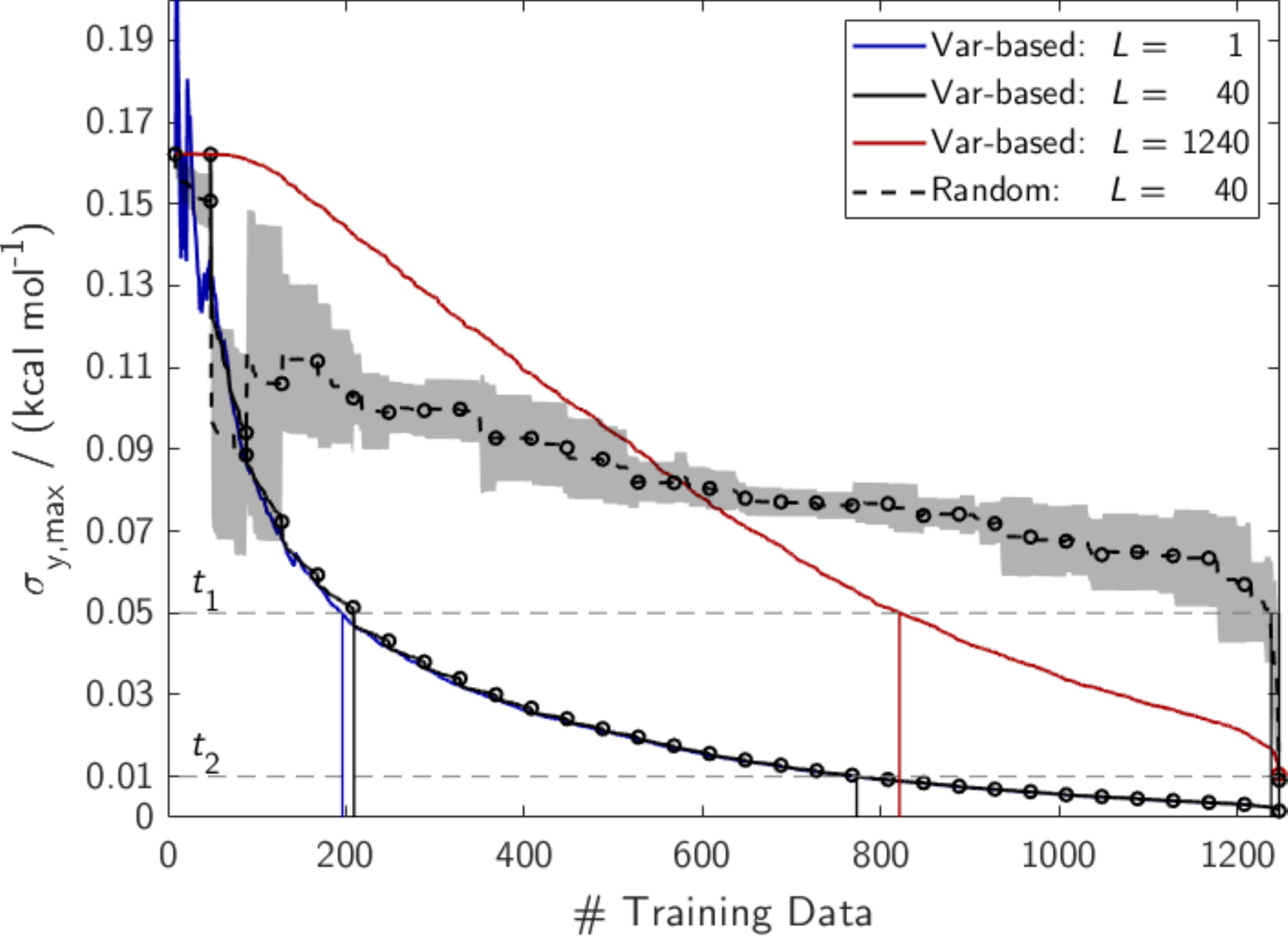}
    \caption{
    The maximum prediction uncertainty as a function of the number of training data.
    The former is measured as the maximum standard deviation of the GP posterior distribution, $\hat{\sigma}_{y,\text{max}}$.
    The initial training set consisted of 8 dimers, which were drawn randomly from the overall data set (\textsf{S13x8}~$+$ \textsf{ROTA}~$+$ \textsf{CONF}).
    The remaining 1,240 dimers served as pool of potential training data (query set).
    The maximum prediction uncertainty relates to the current query set.
    Two sampling strategies were considered,
    BVS with three different batch sizes (${L = 1}$, blue line; ${L = 40}$, solid black line; and ${L =}$\,1,240, red line) and random sampling with a batch size of ${L = 40}$ (black dashed line and gray band obtained from 5 repeated draws).
    Intersections with two possible accuracy thresholds, $t_1$ and $t_2$ (gray dashed lines), are highlighted by vertical lines.
    We employed the Matérn-\sfrac{1}{2} kernel and the \textsf{histD3(BJ)} featurization to produce this figure.
    }
    \label{fig:maxvar}
\end{figure}

    We considered an initial training set $\mathcal{T}_{8}^{8}$ comprising eight randomly drawn target values and the corresponding feature vectors.
    This set was randomly drawn from the overall data set of 1,248 dimers (\textsf{S13x8}~+ \textsf{ROTA}~+ \textsf{CONF}).
    The set of all 1,240 remaining dimers, the query set $\mathcal{Q}_{1240}$, incorporates only the associated feature vectors.
    In a practical setup, electronic-structure calculations would not yet be available for the query set.
    After GP regression on $\mathcal{T}_{8}^{8}$ including hyperparameter optimization, the GP posterior distribution was determined for all feature vectors in $\mathcal{Q}_{1240}$ (query vectors).
    We chose three different batch sizes (${L =}$ 1, 40, 1,240), to demonstrate the advantages and limitations of batch learning (cf.~Section~\ref{sec:var-based}) compared to sequential learning\cite{simm2018} (${L = 1}$).
    In Figure~\ref{fig:maxvar}, we compare BVS against random sampling by examining the maximum posterior standard deviation, $\hat{\sigma}_{y,\text{max}}$, of the query set as a function of the number of training data.
    The maximum standard deviation can be interpreted as the maximum prediction uncertainty.
    To cover the entire range of training data, we set the accuracy threshold $t$ to zero.
    The results shown correspond to the Matérn-\sfrac{1}{2} kernel and the \textsf{histD3(BJ)} featurization.
    
    Initially, $\hat{\sigma}_{y,\text{max}}$ fluctuates strongly in the case of variance-based sampling with sequential learning (${L = 1}$, blue line).
    As the effect of a new data point on the GP posterior distribution is larger for small training set sizes, the hyperparameters are more sensitive to new data points in the initial phase.
    This hyperparameter sensitivity explains the observed fluctuations.
    After adding about 50 training data points, $\hat{\sigma}_{y,\text{max}}$ decays rapidly, indicating a rapid improvement of the prediction accuracy with respect to the query set.
    The profile of $\hat{\sigma}_{y,\text{max}}$ for BVS with a batch size of ${L = 40}$ (black solid line) is very similar to the ${L = 1}$ case, but does not exhibit fluctuations in the initial phase.
    This apparently more stable behavior can be explained by the initially fixed hyperparameters, which are only reoptimized after the addition of 40 query vectors to the training set.
    As $t = 0$, all following hyperparameter optimizations (indicated by the black circles) are also performed after selecting batches of 40 query vectors.
    In a practical setup, this approach would be 40 times more efficient than the ${L = 1}$ approach, without significantly altering the number of training data points for a given threshold beyond 50 training data points.
    For the hypothetical thresholds ${t_1 = 0.05}$~kcal~mol$^{-1}$ and ${t_2 = 0.01}$~kcal~mol$^{-1}$ (gray dashed lines), the algorithm would stop at about 200 and 770 training data points, respectively, in both cases (${L = 1}$ and ${L = 40}$).
    
    The profile of $\hat{\sigma}_{y,\text{max}}$ will change significantly if the hyperparameters are not updated while adding training data points as in the case of BVS with a batch size of ${L =}$\,1,240 (red line).
    Here, the maximum posterior standard deviation is significantly overestimated throughout.
    For instance, the first hypothetical threshold ${t_1 = 0.05}$~kcal~mol$^{-1}$ is only reached after adding 821 data points to the training set, which would amount to an overhead of about 3,700 DLPNO-CCSD(T) single-point calculations compared to the ${L = 1}$ and ${L = 40}$ cases.
    Hence, the actual batch size employed in BVS is an important parameter that requires careful adjustment.
    
    Compared to BVS, the decay rate of $\hat{\sigma}_{y,\text{max}}$ with respect to the number of training data is much lower in the case of random sampling (black dashed line and gray band resulting from 5 repetitions; $L = 40$).
    For instace, the first hypothetical threshold ${t_1 = 0.05}$~kcal~mol$^{-1}$ is only reached after adding almost all query vectors (1,238) to the training set.
    As the posterior variance is not taken into account, potentially many uninformative (low-variance) query vectors are drawn, which renders random sampling highly inefficient for the generation of small informative training sets.
    
    It seems that BVS with a batch size of ${L = 40}$ is the most efficient option among the ones considered.
    However, the posterior variance is only an estimate of the scatter of the prediction error and not the prediction error itself.
    As the GP posterior distribution is Gaussian, 95.5\% of all D3(BJ)-GP errors should be smaller than $\hat{\sigma}_{y,\text{max}}$.
    We find that the Matérn-\sfrac{1}{2} kernel fulfills this condition after the addition of the first and all following batches.
    While the Matérn-\sfrac{3}{2} yields slightly higher decay rates of $\hat{\sigma}_{y,\text{max}}$ (Figure~S3), it violates this condition in an unpredictable manner.
    As the statistical validity of the posterior variance is of utmost importance for the efficient and reliable generation of informative training sets, we choose to incorporate the Matérn-\sfrac{1}{2} kernel into the D3(BJ)-GP model.
    
\begin{figure} [!ht]
    \centering
    \includegraphics[width=0.6\textwidth]{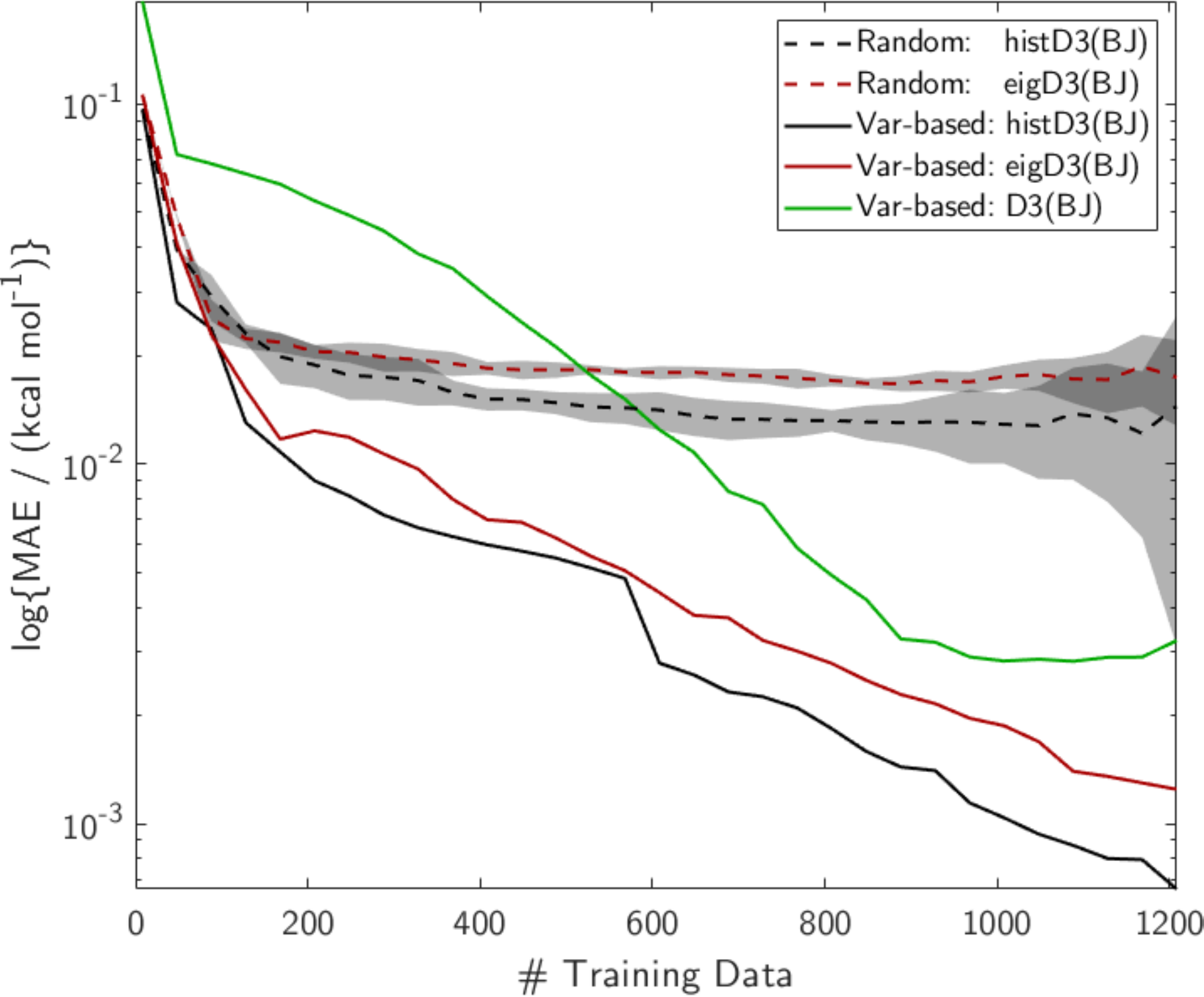}
    \caption{
    The prediction error (as measured by the logarithm of the MAE) of the D3(BJ)-GP model for PBE as a function of the number of training data.
    The initial training set consisted of 8 dimers, which were drawn randomly from the overall data set (\textsf{S13x8}~$+$ \textsf{ROTA}~$+$ \textsf{CONF}).
    The remaining 1,240 dimers served as pool of potential training data (query set).
    The prediction error relates to the current query set.
    Two sampling strategies were considered,
    BVS (black and red solid lines) and random sampling (dashed lines and gray bands obtained from 5 repeated draws).
    In all cases, we used a constant batch size of ${L = 40}$.
    Additionally, we considered the prediction error of the D3(BJ) model for PBE.
    We employed the Matérn-\sfrac{1}{2} kernel and both \textsf{histD3(BJ)} (black lines) and \textsf{eigD3(BJ)} (red lines) featurizations to produce this figure.
    }
    \label{fig:logmae}
\end{figure}

    The actual prediction error of the D3(BJ)-GP model for PBE obtained from both BVS and random sampling is shown in Figure~\ref{fig:logmae}.
    It is based on the same initial training and query sets as those employed for the production of Figure~\ref{fig:maxvar}.
    In all cases, a batch size of ${L = 40}$ was chosen.
    We employed the Matérn-\sfrac{1}{2} kernel and both featurizations introduced in this work, \textsf{eigD3(BJ)} (black lines) and \textsf{histD3(BJ)} (red lines).
    Mean (dashed lines) and standard deviation (gray bands) of the prediction errors obtained via random sampling are the result of 5 repeated draws.
    Additionally, we report the prediction error of the D3(BJ) model for PBE (green line).
    Note that we consider the MAE instead of the MARE as the posterior variance is learned from \textit{absolute} residual interaction energies.
    Learning their relative analog is inefficient as the division by the reference interaction energy would annihilate the smoothness of the target function gained by applying the $\Delta$-learning approach\cite{ramakrishnan2015} introduced by von Lilienfeld and co-workers.

    BVS leads to significantly smaller prediction errors than random sampling, independent of the number of training data considered.
    This finding confirms the hypothesis that query data associated with a high posterior variance likely contribute to the overall error in a significant way.
    For both sampling strategies, we find that the \textsf{histD3(BJ)} featurization leads to systematically smaller prediction errors than its \textsf{eigD3(BJ)} analog, which is in agreement with the findings discussed in Section~\ref{sec:learning-curves}.
    Note that the query sets obtained with random sampling and BVS are different and, therefore, a comparison of the corresponding prediction errors is somewhat biased.
    For this reason, we only refer to the prediction errors associated with BVS in the following.
    
    As the query set decreases constantly, two effects are intermingled in the D3(BJ)-GP prediction error:
    The degree of improvement by adding more training data (learning rate) and the improvement by removing instances with large absolute errors from the query set.
    The change of the D3(BJ) prediction error supports unfolding these effects as it is not affected by the training data (it would be constant for a fixed test set).
    Hence, the change in the difference between the D3(BJ)-GP and D3(BJ) prediction errors can be interpreted as the learning rate of the D3(BJ)-GP model.
    It only increases until the addition of about 200 training data points (and again for $>$1,000 training data points).
    However, the learning rate is positive throughout as the D3(BJ)-GP prediction error is constantly smaller than its D3(BJ) complement.
    This finding is independent of the featurization.
    
\section{Conclusions}

    We introduced the D3(BJ)-GP model, which adjusts for systematic errors in D3(BJ)-type dispersion corrections by harnessing the capabilities of Gaussian process (GP) regression.
    The model learns mappings from linear combinations of atom-pairwise D3(BJ) terms (input) to errors in PBE-D3(BJ)/ma-def2-QZVPP interaction energies (target) with respect to DLPNO-CCSD(T)/CBS.
    We generated input--target pairs for 1,248 molecular dimers, which resemble the dispersion-dominated systems contained in the S66x8 data set.\cite{rezac2008, rezac2011}
    Our systems represent various out-of-equilibrium distances, conformations, and relative orientations.
    As the empirical D3(BJ) parameters ($a_1$, $a_2$, $s_8$) were previously optimized with respect to CCSD(T)/CBS interaction energies, we performed a reparametrization of the D3(BJ) model for PBE with respect to 66 dimers selected from our data set.
    At the global minimum, $a_1$ and $s_8$ equal zero and $a_2 = 5.6841$~bohr, effectively eliminating all dipole--quadrupole terms from the D3(BJ) energy expression.
    We found that these three parameters perform best for the remaining 1,182 dispersion-dominated dimers compared to previously published PBE-specific D3(BJ) parameter sets.\cite{grimme2011, smith2016, weymuth2018}
    When repeating the optimization after replacing the DLPNO-CCSD(T)/CBS interaction energies by their CCSD(T)/CBS analogs, the global minimum was still located at $a_1$ and $s_8$ equal to zero.
    This finding suggests that a single empirical D3(BJ) parameter, $a_2$, is sufficient to yield reliable dispersion corrections for dispersion-dominated systems, at least for the PBE functional.
    
    Compared to the D3(BJ) model, which has a fixed functional form and is global with respect to its parameters, the D3(BJ)-GP model has a high degree of functional flexibility as it is based on a local, nonlinear kernel.
    Therefore, D3(BJ)-GP-type dispersion corrections can be significantly improved for a given domain of application by adding new input--target pairs to the training set representing that domain.
    Regarding interpolation, we showed that the D3(BJ)-GP model is associated with a steep learning curve and starts to improve over D3(BJ) after training it on less than five data points.
    After the addition of about 30 data points, the predictions continued to improve only marginally.
    Regarding extrapolation, the accuracy of dispersion corrections improved only slightly (by about 1~\%) or even declined with a close-to-zero or even negative learning rate.
    These results show that the performance improvement of the D3(BJ)-GP model critically depends on the composition of the training set.
    It is not enough to sample the domain of application uniformly as we do not know the number of data points required to achieve a certain accuracy beforehand.
    Furthermore, uniform sampling does not distinguish between relevant and irrelevant input dimensions, which are also unknown beforehand.
    Hence, selecting the training set a priori, as done in the case of the original D3(BJ) model, is a poor strategy for learning D3(BJ)-GP-type dispersion corrections.

    To optimally increase the performance of the D3(BJ)-GP model, it is built on dynamic training sets.
    We aimed at selecting training sets that are maximally informative, i.e., that contain the least number of data points to achieve a predefined accuracy over a given domain of application.
    For this purpose, we exploited the variance of the GP posterior distribution.
    Of all potential training systems considered, the system with the highest posterior variance is selected (referred to as variance-based sampling).
    If interaction energies are not yet available for the new training system, electronic-structure calculations must be carried out.
    Subsequently, the GP posterior distribution is updated and the next training system is selected.
    Through this active learning process, we obtain a system-focused, self-improving model for dispersion interactions equipped with confidence intervals.
    We demonstrated that the posterior variance can be approximately updated from only the input variables of the new training system, which are obtained efficiently from D3(BJ) calculations.
    This way, we collect a batch of new training systems before the corresponding electronic-structure calculations are being carried out at the same time.
    We refer to this selection approach as batch-wise variance-based sampling (BVS).
    BVS-guided active learning renders our D3(BJ)-GP workflow efficient \textit{and} controllable.
    We showed that dynamic training sets significantly improve the learning rates of the D3(BJ)-GP model.
    For the data set studied, it systematically yields improved dispersion corrections, which can be calculated almost as efficiently as their semiclassical D3(BJ) complements. 
    Like its predecessor D3, the D3-GP approach is implemented in a black-box fashion, the difference being that the latter requires a specification of the training set (representing the domain of application).
    Once benchmark data are provided for the associated systems (we recommend an automated selection by our BVS-guided active learning scheme), the D3(BJ)-GP model will update itself through an optimization of the underlying GP hyperparameters.
    Note that the D3 parameters (here, of the BJ damping function) are not affected by this procedure.
    They are only determined once as in the original D3 approach.
    
    Although we have examined the D3-GP approach with respect to the PBE functional and the BJ damping function only, the D3-GP workflow can be straightforwardly generalized without further effort as it is independent of the electronic-structure approximation and the damping scheme.
    As it is also independent of the dispersion coefficient model, the D3-GP workflow would essentially be the same for the D4 approach.
    Moreover, modifications of the input variables (molecular representations) will render any D$x$-GP approach also applicable for other, non-D$x$-type dispersion corrections.
    One could also learn reference interaction energies directly\,---\,instead of indirectly via the residual interaction energy defined in Eq.~(\ref{eq:residual}).
    In that case, however, the physical information already encoded in the quantity to be corrected (here, PBE-D3(BJ)/ma-def2-QZVPP interaction energies) would need to be encoded into the input variables, which is an intricate and ambiguous task.
    Eventually, the workflow is not even problem-dependent.
    Our BVS-guided active learning scheme combined with GP regression could be helpful for a diverse range of prediction tasks.
        
    To make our D$x$-GP approach generally accessible, we will release a version-controlled database in future work.
    This database can then be used and updated to collectively improve on the D$x$-GP model.
    Version control is crucial as the database is dynamic, which affects the composition of the training set generated by BVS-guided active learning.
    With version control, previous results can be permanently reproduced even though the database has been altered meanwhile.

\section*{Acknowledgments}

    J.P.~acknowledges funding through an \textit{Early Postdoc.Mobility} fellowship by the Swiss National Science Foundation (project~no.\,178463).
	S.G.~and M.R.~are grateful for financial support by the Swiss National Science Foundation (project~no.\,200021\_182400).
	

\providecommand{\refin}[1]{\\ \textbf{Referenced in:} #1}



\end{document}